\documentclass{ws-ijmpa}
\usepackage[super,compress]{cite}
\usepackage{graphicx}
\begin{document}
\markboth{Yabing ZUO, Yue HU, Linlin HE, Wei YANG, Yan CHEN, Yannan
HAO}{$D \rightarrow a_1, f_1$ transition form factors and
semileptonic decays via 3-point QCD sum rules}

%
\catchline{}{}{}{}{}
%

\title{$D \rightarrow a_1, f_1$ transition form factors and semileptonic decays via 3-point QCD sum rules
}

\author{Yabing ZUO\footnote{zuoyabing@lnnu.edu.cn
}, \hspace{0.1cm} Yue HU\footnote{1152332628@qq.com}, \hspace{0.1cm}
Linlin HE\footnote{748254980@qq.com}, \hspace{0.1cm} Wei
YANG\footnote{974049627@qq.com}, \hspace{0.1cm} Yan
CHEN\footnote{1031632684@qq.com}, \hspace{0.1cm} Yannan
HAO\footnote{2458067731@qq.com}}

\address{School of Physics and Electronic Technology, Liaoning Normal University\\
 Dalian 116029,  P.R.China
}



\maketitle

\begin{history}
\received{Day Month Year}
\revised{Day Month Year}
\end{history}

\begin{abstract}

By using the 3-point QCD sum rules, we calculate the transition form
factors of $D$ decays into the spin triplet axial vector mesons
$a_1(1260)$, $f_1(1285) $, $f_1(1420)$. In the calculations, we
consider the quark contents of each meson in detail. In view of the
fact that the isospin of $a_1(1260)$ is one, we calculate the $D^+
\rightarrow a_1^0 (1260)$ and $D^0 \rightarrow a_1^- (1260)$
transition form factors separately. In the case of $ f_1(1285),
f_1(1420)$, the mixing between light flavor $SU(3)$ singlet and
octet is taken into account. Based on the form factors obtained
here, we give predictions for the branching ratios of relevant
semileptonic decays, which can be tested in the future experiments.

\end{abstract}

\ccode{PACS numbers:13.20.-v, 13.20.Fc }


\section{Introduction}

The heavy to light meson exclusive decays are very important in
testing the particle physics standard model (SM), extracting its
parameters and searching for possible new physics. So far, the
decays with only S-wave mesons (pseudoscalar mesons, vector mesons
etc.) in the final states have been analyzed extensively both from
theoretical and experimental aspects. Comparatively speaking, few
studies have been done on those decays involving P-wave mesons
(scalar mesons, axial vector mesons etc.) in the final states. In
the last ten years, with the development of experiments, a large
amount of decays with P-wave mesons in the final states have been
found and measured\cite{PDG, BA1,BA2,BA3,BA4,BA5, KA}, which promote
the theoretical investigations on these decays. To give predictions
for the exclusive decays, one needs the knowledge of transition form
factors. K.C. Yang et al. gave the form factors of $B_{(s)}$ to
P-wave mesons by using the light cone sum rules and studied relevant
exclusive decays\cite{KCY1,KCY2,KCY3}.

A special role is played by the $D_{(s)}$ exclusive decays, which
provide an unique place to probe the new physics coupling with up
type quarks. However, the light cone sum rules adopted in
Ref.\cite{KCY1,KCY2,KCY3} are based on an expansion over $m_A/m_b$
($m_A$ and $m_b$ denote the mass of axial vector meson and bottom
quark respectively) and therefore can not be used to calculate the
form factors of $D$ to axial vector mesons. R. Khosravi et al.
calculated the $D_{(s)}$ to $K_1$ transition form factors in 3-point
QCD sum rules\cite{RK}. In the present paper, we intend to calculate
the semileptonic form factors of $D$ to spin triplet axial vector
mesons $a_1(1260)$, $f_1(1285) $, $f_1(1420)$ by using the 3-point
QCD sum rules and give a prediction for the branching ratios of
relevant semileptoinc decays.

\section{Definitions and expressions of $D$ to axial vector meson transition form factors via 3-point QCD sum rules}

Following Ref.\cite{RK}, the semileptonic $D$ to axial vector meson
$A$ transition form factors are defined via corresponding hadronic
matrix elements as follows,
\begin{eqnarray}
&&  \langle A(p^\prime, \varepsilon ) |\bar{d} \gamma_\mu \gamma_5 c
| D (p) \rangle  = - \frac{ 2 f_V (q^2) }{ (m_D +m_A)} \epsilon_{\mu
\nu \alpha \beta} \varepsilon^\nu p^\alpha p^{\prime \beta } \\
&&  \langle A ( p^\prime, \varepsilon ) | \bar{d} \gamma_\mu c | D
(p) \rangle = i \left [ f_0(q^2)(m_D +m_A ) \varepsilon_\mu  -
\frac{f_1(q^2)}{( m_D + m_A )} \varepsilon \cdot p P_\mu \right. \nonumber \\
&& \hspace{3.5cm}\left.- \frac{ f_2 (q^2)}{(m_D +m_A)} \varepsilon
\cdot p q_\mu \right ]
\end{eqnarray}
Here $A$ denotes the axial vector meson $a_1$ or $f_1$. $P_\mu =
(p+p^\prime)_\mu$, $q_\mu = (p-p^\prime)_\mu$ and $\varepsilon$ is
the polarization vector of $A$.

The form factors $f_V$, $f_0$, $f_1$, $f_2$ can be evaluated via
3-point QCD sum rules. As stated in Ref.\cite{RK}, the following
redefinitions are needed in order for the calculations to be simple,
\begin{eqnarray}
& & F_V (q^2)= \frac{2 f_V (q^2)}{m_D+m_A}, \hspace{0.5cm} F_0 (q^2)
=
(m_D +m_A) f_0 (q^2) \nonumber \\
& & F_1(q^2) = - \frac{f_1(q^2)}{m_D +m_A}, \hspace{0.5cm} F_2(q^2)
= - \frac{f_2(q^2)}{m_D+m_A}
\end{eqnarray}
Using the same analysis as Ref.\cite{RK} with simply choosing $D_q$
to be $D$ and replacing $K_{1A(B)}$ with $A$, we have
\begin{eqnarray}\label{sumrule}
& & F_i ( q^2, s_0, s^\prime_0, M^2_1, M^2_2 ) = - \frac{(m_c +
m_q)}{f_D m^2_D f_A m_A} e^{(m^2_D/M^2_1)}  e^{ (m^2_A/M^2_2)} \left
\{ -\frac{1}{4 \pi^2} \int^{s^\prime_0}_{m^2_c} d s^\prime \right.
\nonumber
\\
& & \hspace{3.5cm} \times \int ^{s_0}_{s_L} d s \rho_i (s, s^\prime,
q^2 ) e^{(-s/M^2_1)} e^{(-s^\prime/M^2_2)} \nonumber \\
& & \hspace{3.5cm} + M^2_1 M^2_2 {\cal B}_{p^2} (M^2_1) {\cal
B}_{p^{\prime 2}} ( M^2_2) \Pi^{\rm non-per}_i ( p^2, p^{\prime 2},
q^2 ) \}
\end{eqnarray}
where $i = V,0,1,2$. The expressions of spetral densities $\rho_i (
s, s^\prime, q^2)$ and corresponding nonperturbative contributions
$\Pi^{\rm non-per}_i (p^2,p^{\prime 2}, q^2 )$ can be found in
Ref.\cite{RK}. $s_0$ and $s^\prime_0$ are the continuum thresholds
in $D$ and axial vector meson $A$ channels respectively. The lower
limit $s_L$ in the integration over $s$ is
\begin{eqnarray}
s_L= \frac{(m^2_q + q^2 -m^2_c -s^\prime )( m^2_c s^\prime - m^2_q
q^2 )}{ ( m^2_c -q^2)( m^2_q - s^\prime )}
\end{eqnarray}

\section{Numerical results and analysis of the form factors}

With the sum rule expressions above, it is now in the position to
evaluate the $D$ to axial vector meson transition form factors.
According to the quark model\cite{PDG}, the flavor contents of
relevant mesons are
\begin{eqnarray}\label{flavor}
& & D^+  = c\bar{d}, \hspace{0.5cm}  D^0  = c \bar{u},
\hspace{0.5cm} a_1^0 (1260) = \frac{1}{\sqrt{2}} ( u \bar{u} - d
\bar{d} ), \hspace{0.5cm} a_1^- (1260)  = \bar{u} d,
\nonumber \\
& &  f_1  = \frac{1}{\sqrt{3}} ( u \bar{u}+ d \bar{d} +s \bar{s}),
\hspace{0.5cm} f_8  = \frac{1}{\sqrt{3}} ( u \bar{u}+ d \bar{d} -2 s
\bar{s})
\end{eqnarray}
Note that the physical $f_1(1285), f_1(1420)$ states are mixtures of
light flavor $SU(3)$ singlet $f_1$ and octet $f_8$, which can be
expressed as
\begin{eqnarray}
 f_1 (1285) = f_1 \cos \theta + f_8 \sin \theta, \hspace{0.5cm} f_1
(1420) = - f_1 \sin \theta + f_8 \cos \theta
\end{eqnarray}
For the mixing angle $\theta$, we choose the value $\theta =
23^\circ$ given by H.Y.Cheng based on phenomenological
analysis\cite{HYC}. The masses and decay constants of relevant
mesons are collected in Table\ref{tab1}. Other input parameters are
quark masses and parameters relating to condensates, which are
listed as
\begin{eqnarray}
& & m_c = 1.27 {\rm GeV }, \hspace{0.5cm} m_d = 0.005 {\rm GeV} ,
\nonumber \\
& &  \langle d \bar{d} \rangle =- 1.38 \times 10^{-2} {\rm GeV}^3,
\hspace{0.5cm} m^2_0 = 0.8 {\rm GeV}^2 \nonumber
\end{eqnarray}

\begin{table}[htbp]
\tbl{ Masses and decay constants of relevant mesons. The masses of
$D^+, D^- , a_1(1260) $ are experimental values given by the
Particle Data Group(PDG)\cite{PDG}, while other quantities are
obtained in the QCD sum rule calculation\cite{KCY4}.}
{\begin{tabular*}{10cm}{@{}ccc@{}} \toprule State  & Mass [GeV] &
Decay
constant $f$ [GeV] \\
\colrule 
$D^+$ &  1.87 & $0.222 $ \\
$D^-$ & 1.86 & 0.222 \\
\hline
$a_1(1260)$ & $1.26 $ & $0.238  $ \\
$f_1 $ & $1.28 $ & $0.245 $ \\
$f_8 $ & $1.29 $ & $0.239 $ \\
\botrule
\end{tabular*} \label{tab1}}
\end{table}

\begin{figure}[htbp]
\includegraphics[width=6cm]{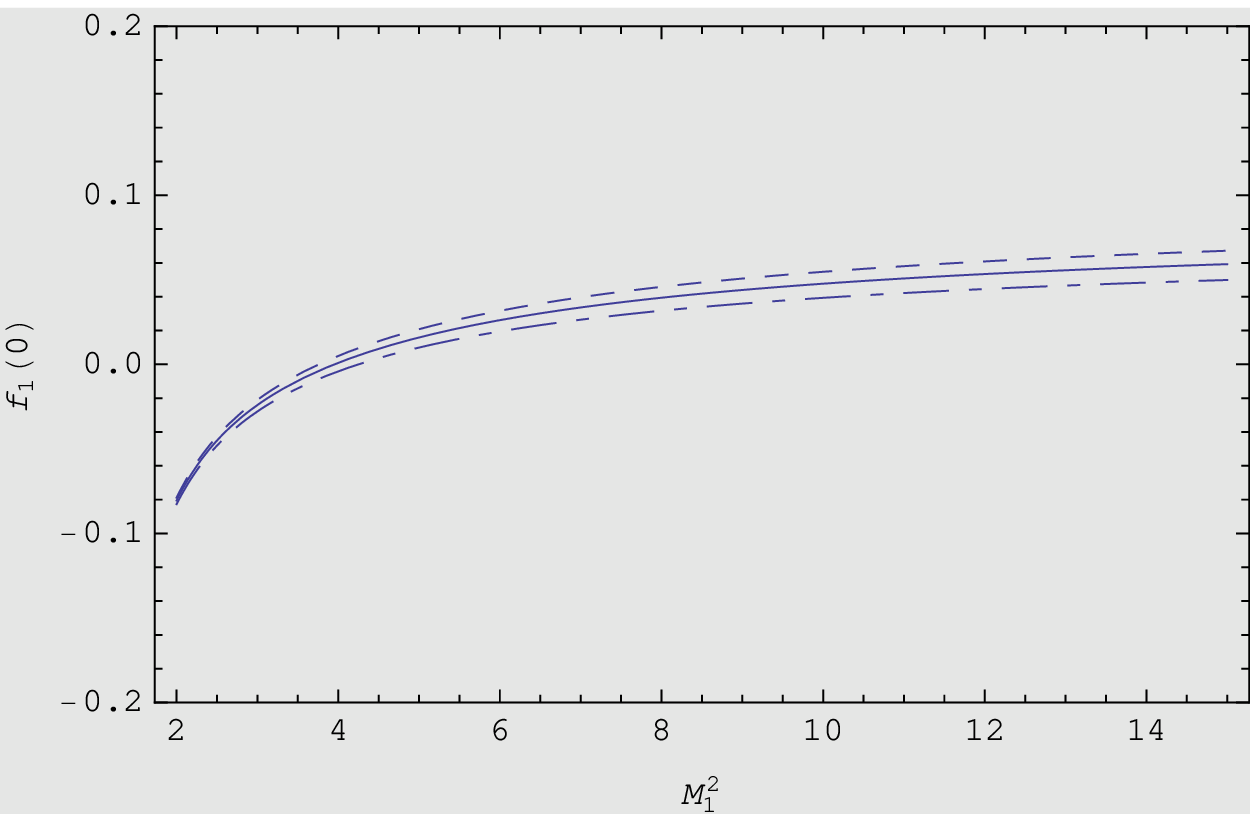}\hfill
\includegraphics[width=6cm]{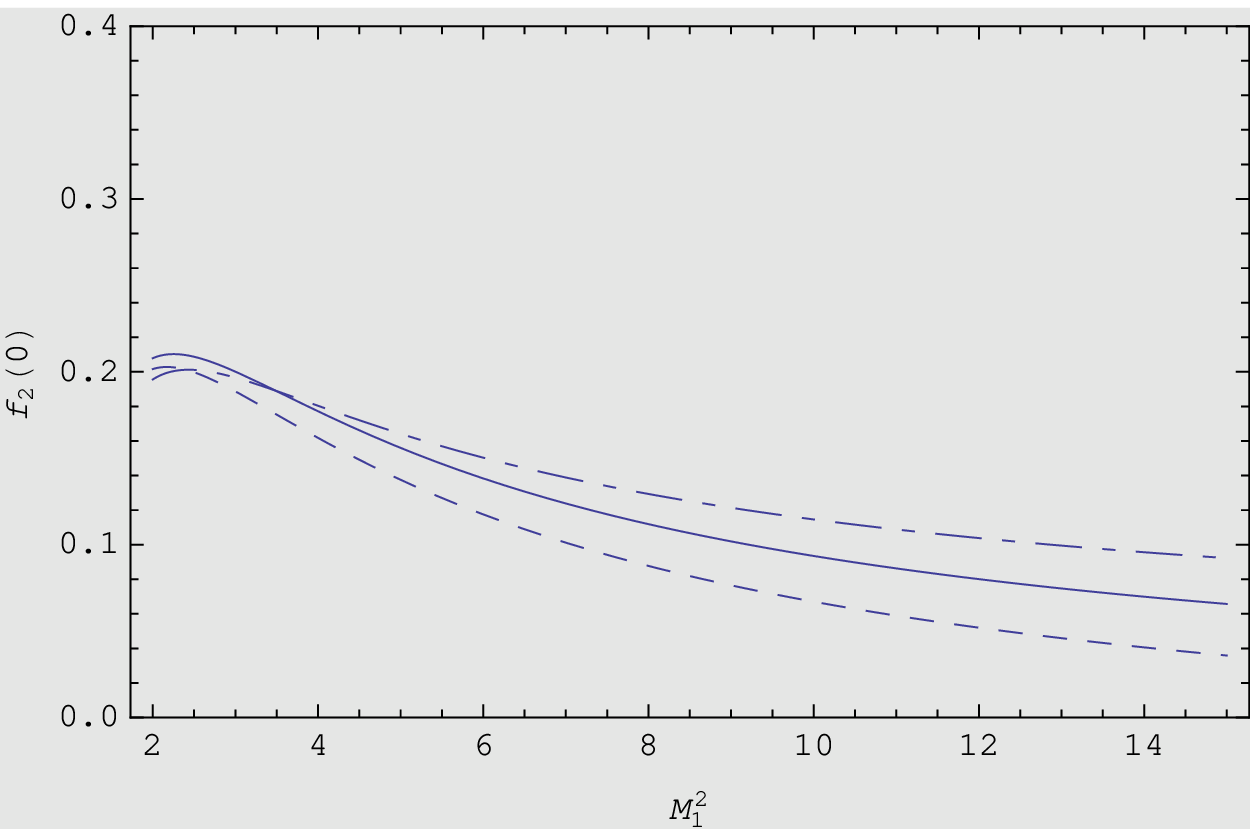}\\
\includegraphics[width=6cm]{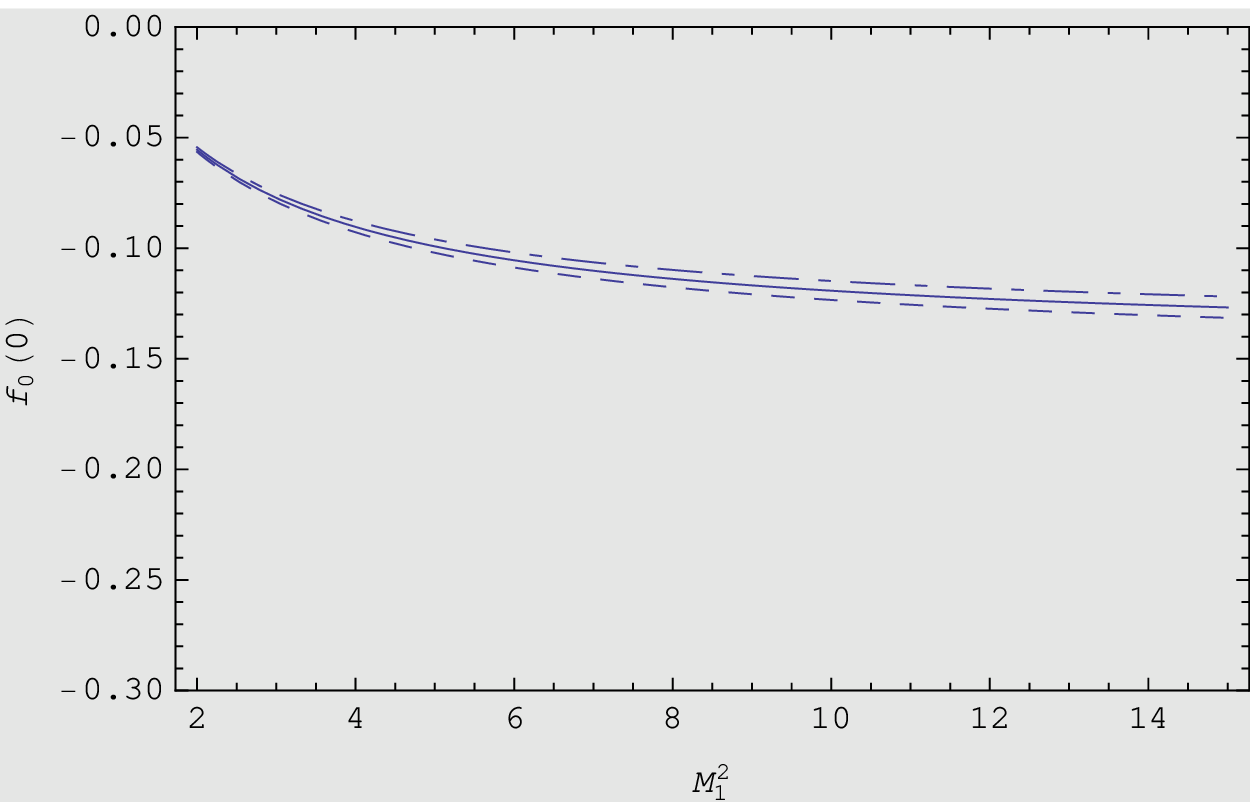}\hfill
\includegraphics[width=6cm]{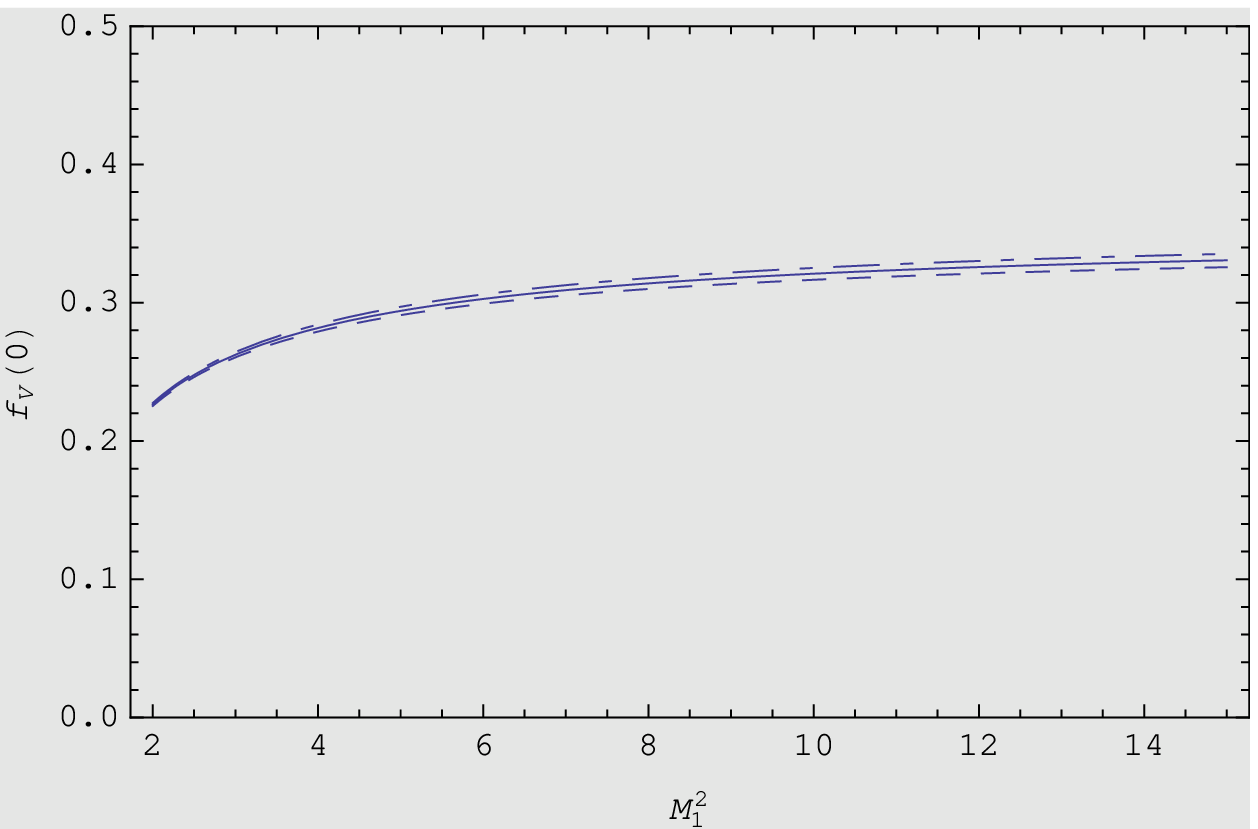}
\caption{$D^+ \rightarrow a_1^0 (1260)$ transition form factors at
$q^2=0$ as functions of $M^2_1$. The dashed, solid and dot dashed
lines correspond to $s_0=7.2$, $7.0$ and $6.8$ ${\rm GeV}^2$
respectively. \label{FM1dpa1}}
\end{figure}
\begin{figure}[htbp]
\includegraphics[width=6cm]{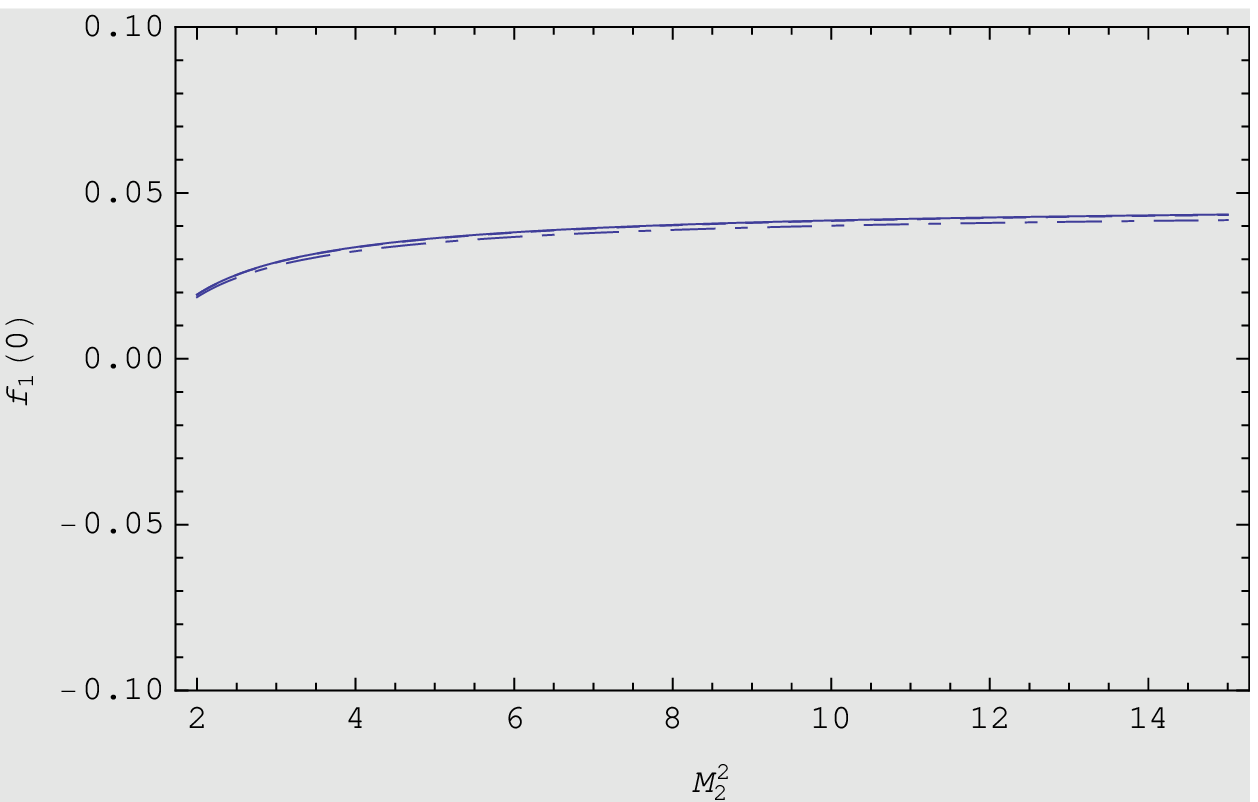}\hfill
\includegraphics[width=6cm]{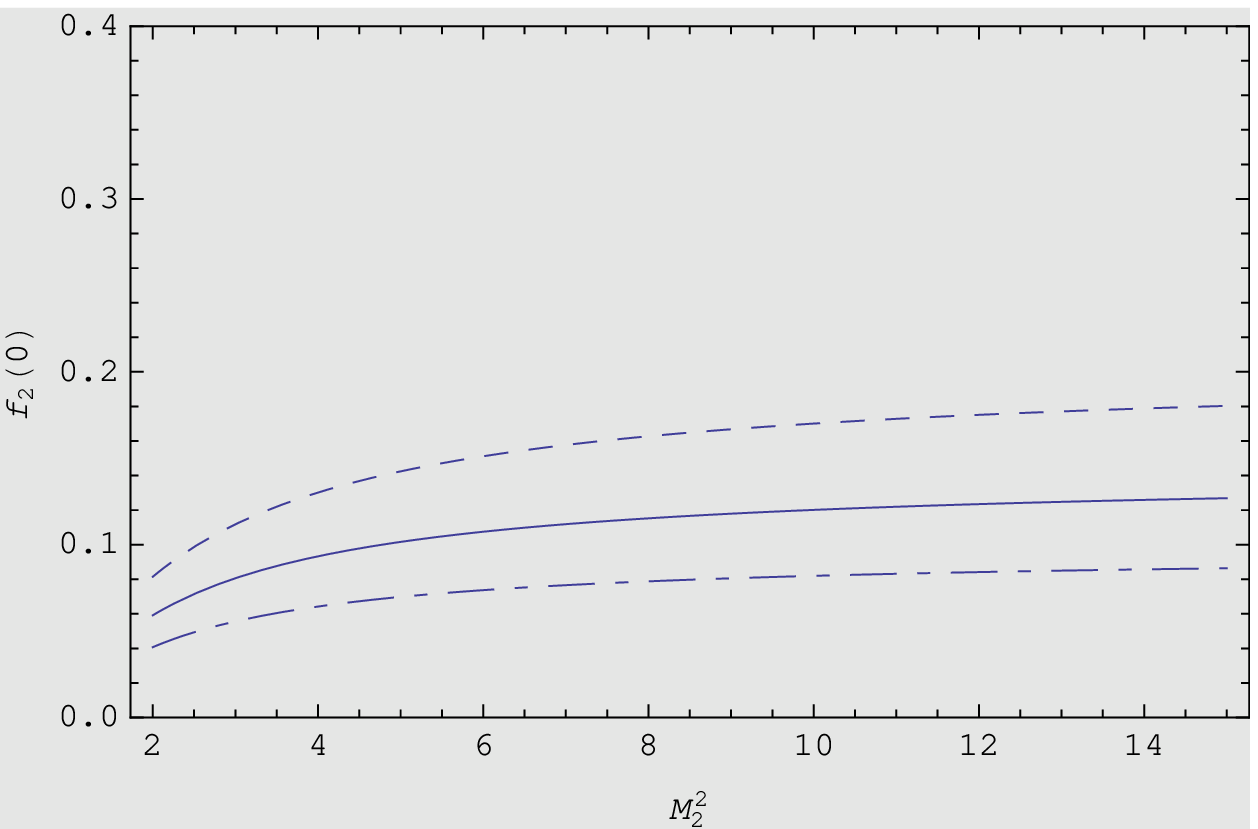}\\
\includegraphics[width=6cm]{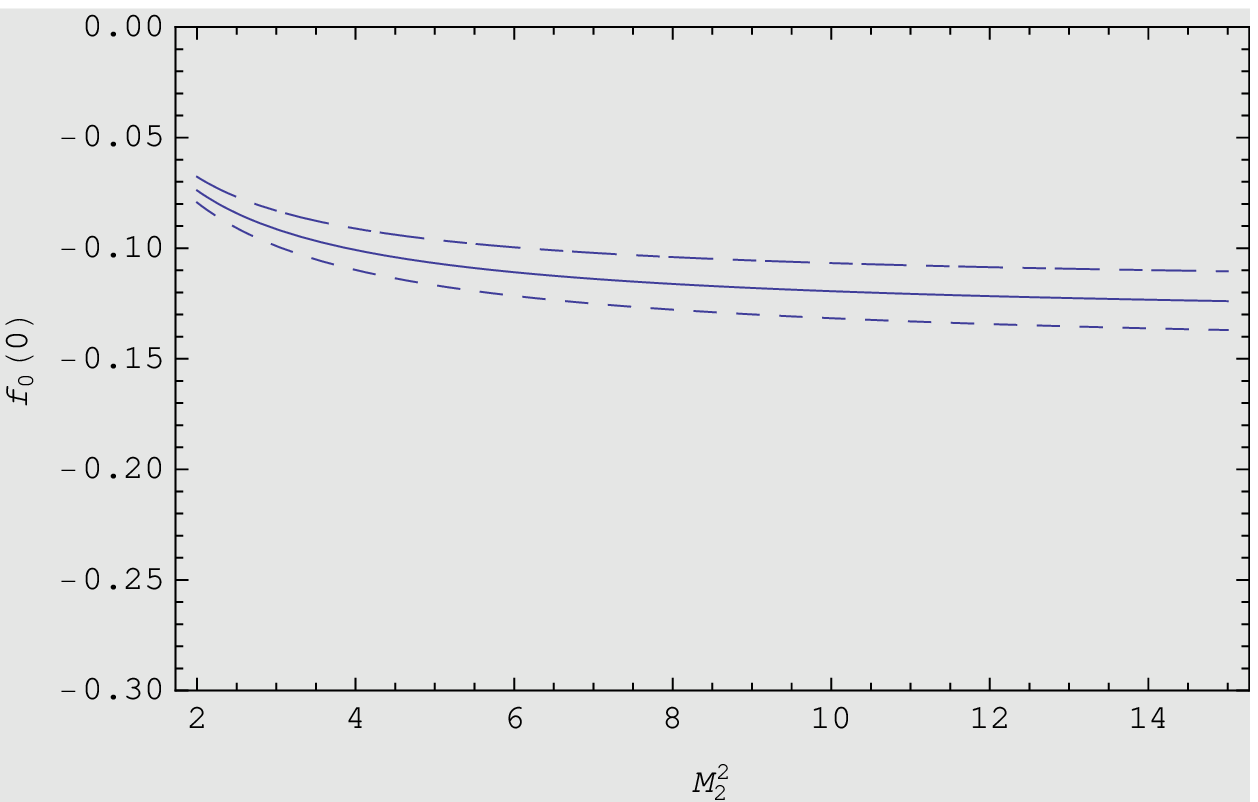}\hfill
\includegraphics[width=6cm]{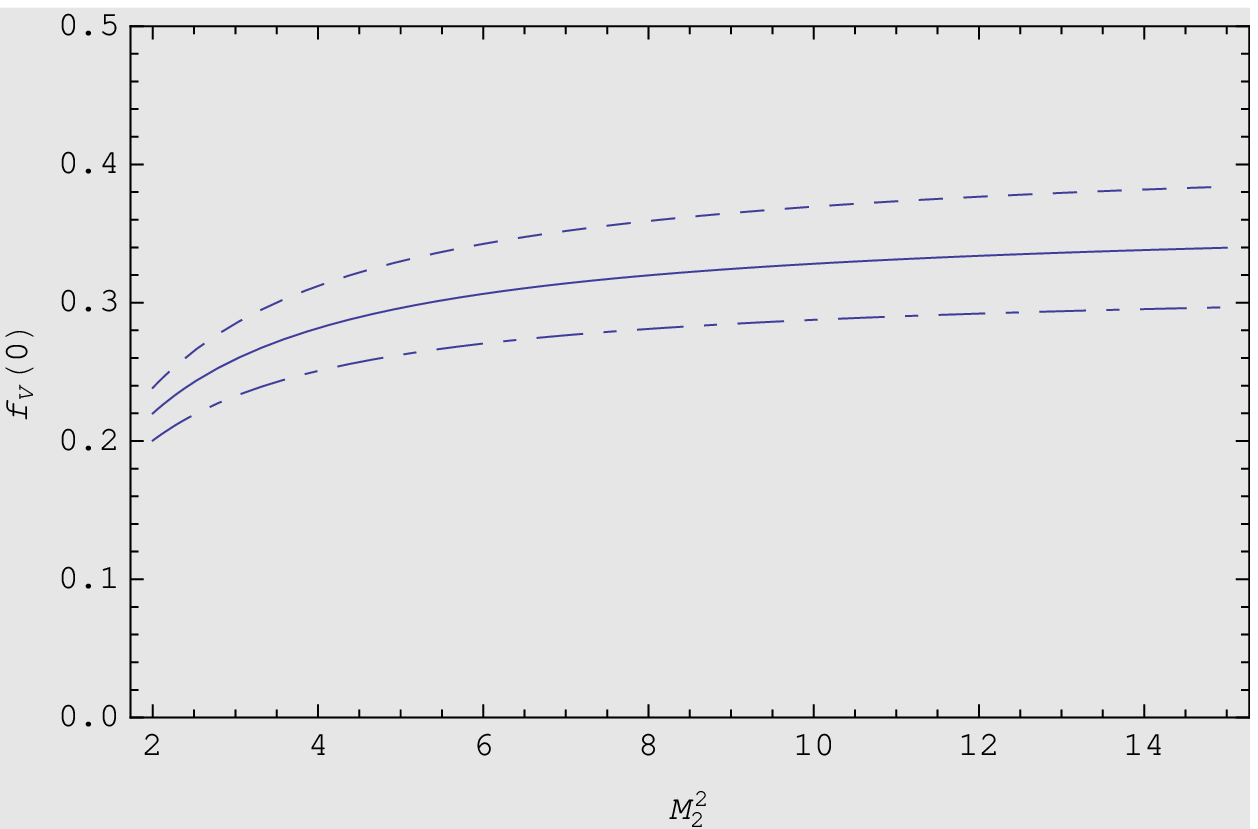}
\caption{ \label{FM2dpa1} $D^+ \rightarrow a_1^0 (1260)$ transition
form factors at $q^2=0$ as functions of $M^2_2$. The dashed, solid
and dot dashed lines correspond to $s_0^\prime=3.7$, $3.5$ and $3.3$
${\rm GeV}^2$ respectively.}
\end{figure}
\begin{figure}[htbp]
\includegraphics[width=6cm]{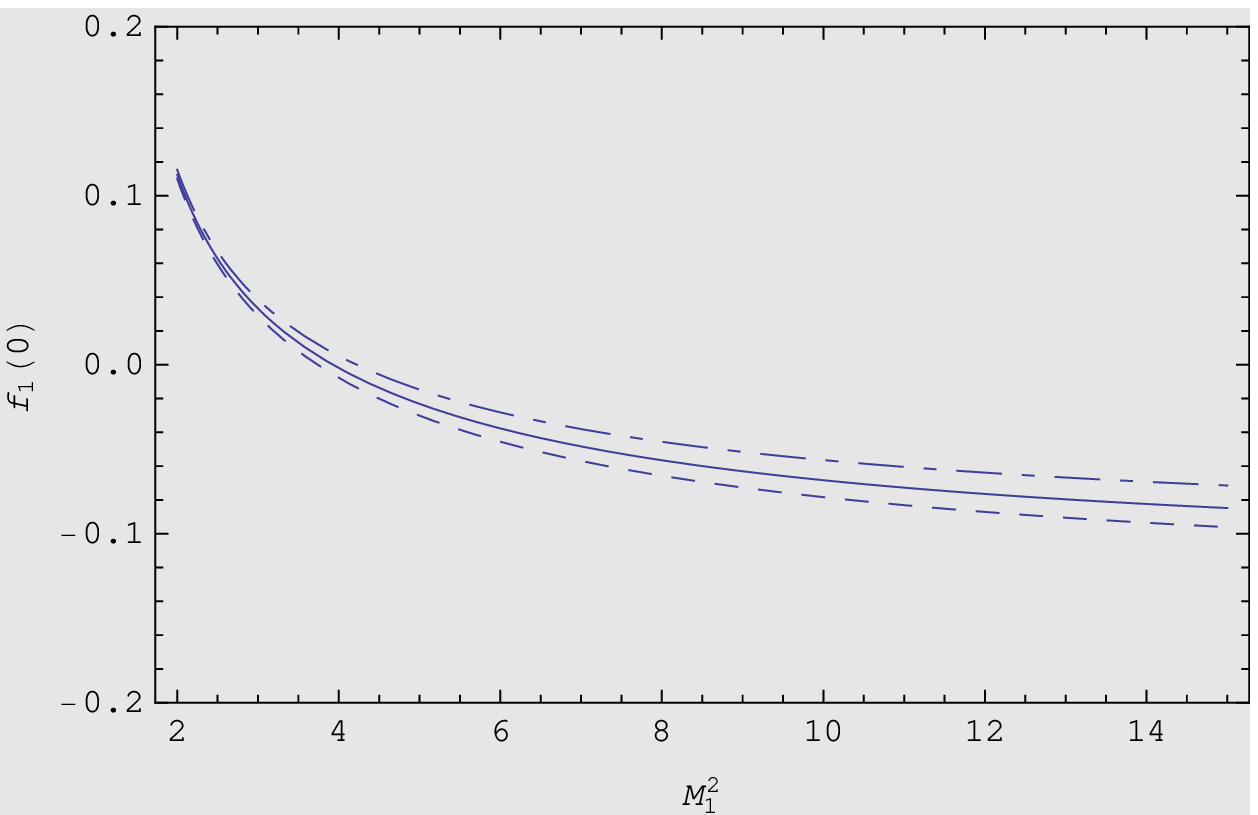}\hfill
\includegraphics[width=6cm]{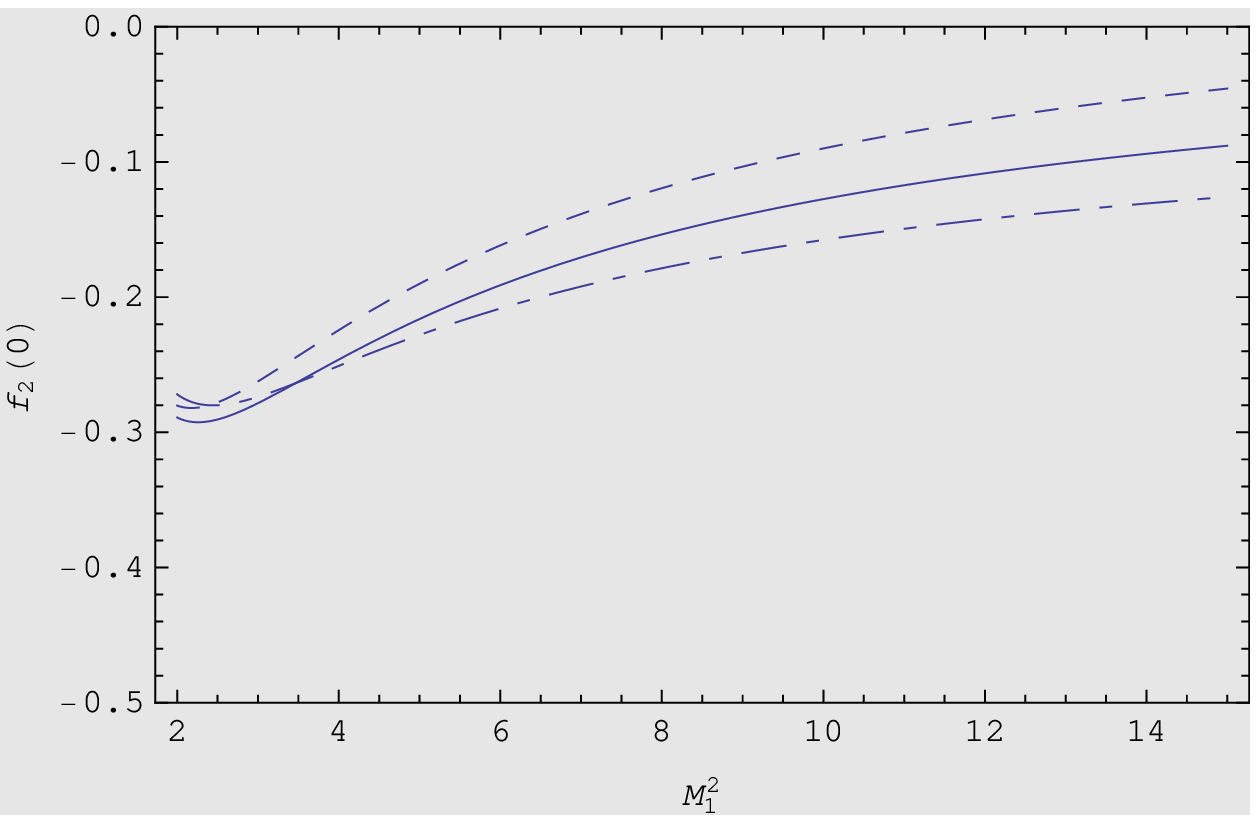}\\
\includegraphics[width=6cm]{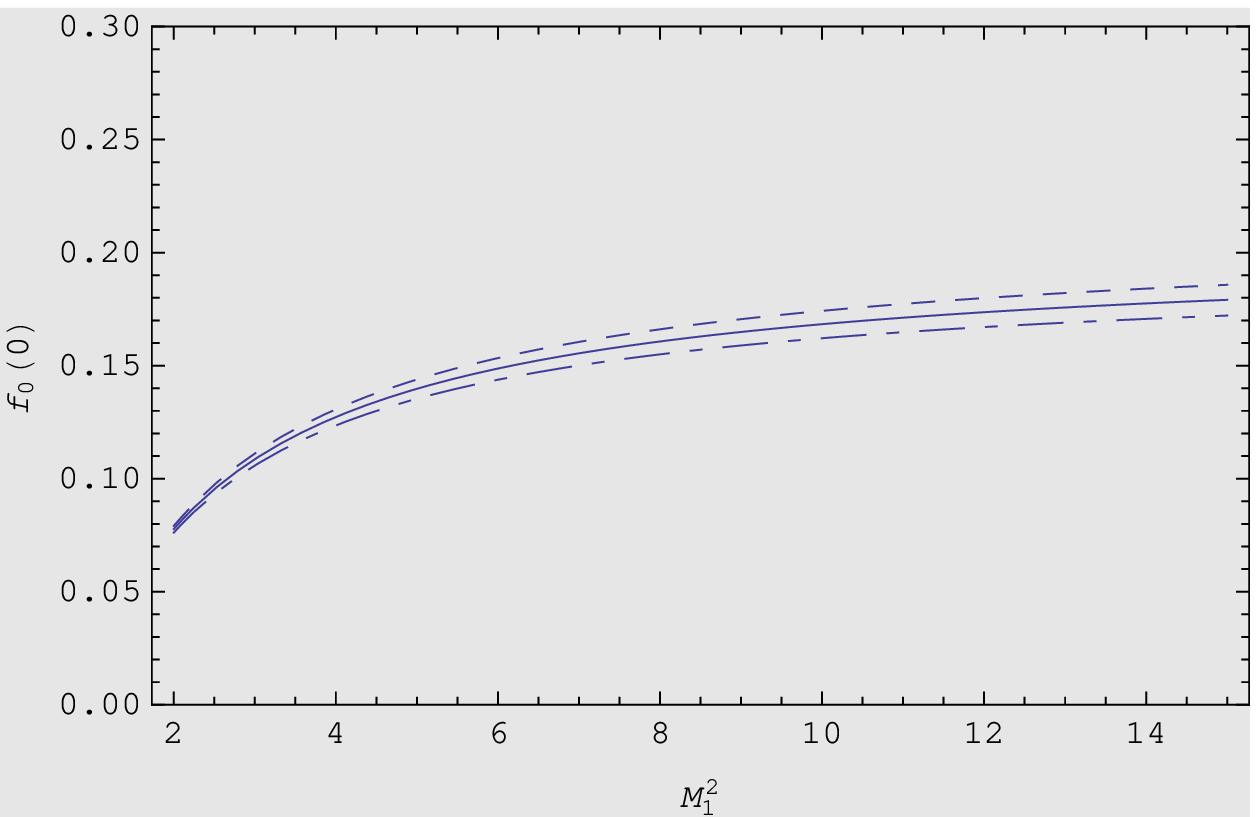}\hfill
\includegraphics[width=6cm]{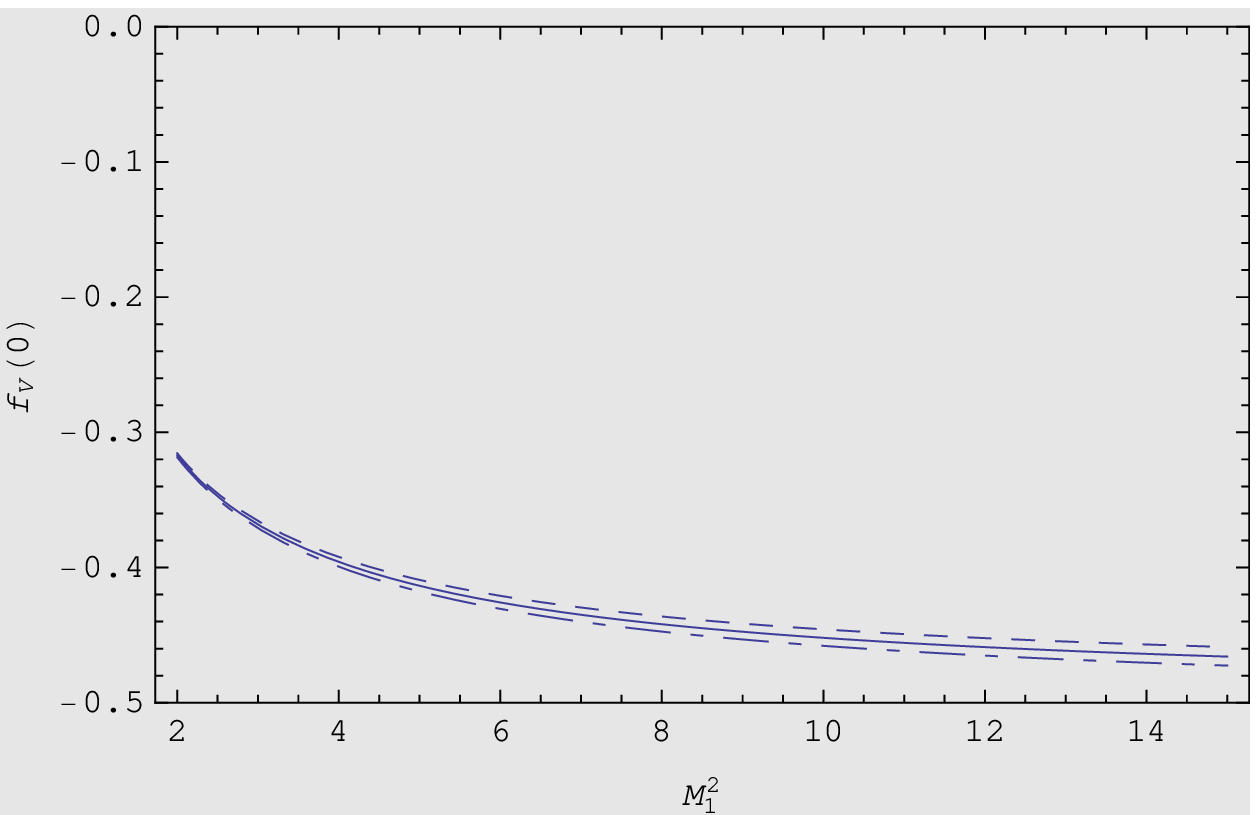}\\
\caption{ \label{FM1d0a1} $D^0 \rightarrow a_1^- (1260)$ transition
form factors at $q^2=0$ as functions of $M^2_1$. The dashed, solid
and dot dashed lines correspond to $s_0=7.2$, $7.0$ and $6.8$ ${\rm
GeV}^2$ respectively.}
\end{figure}
\begin{figure}[htbp]
\includegraphics[width=6cm]{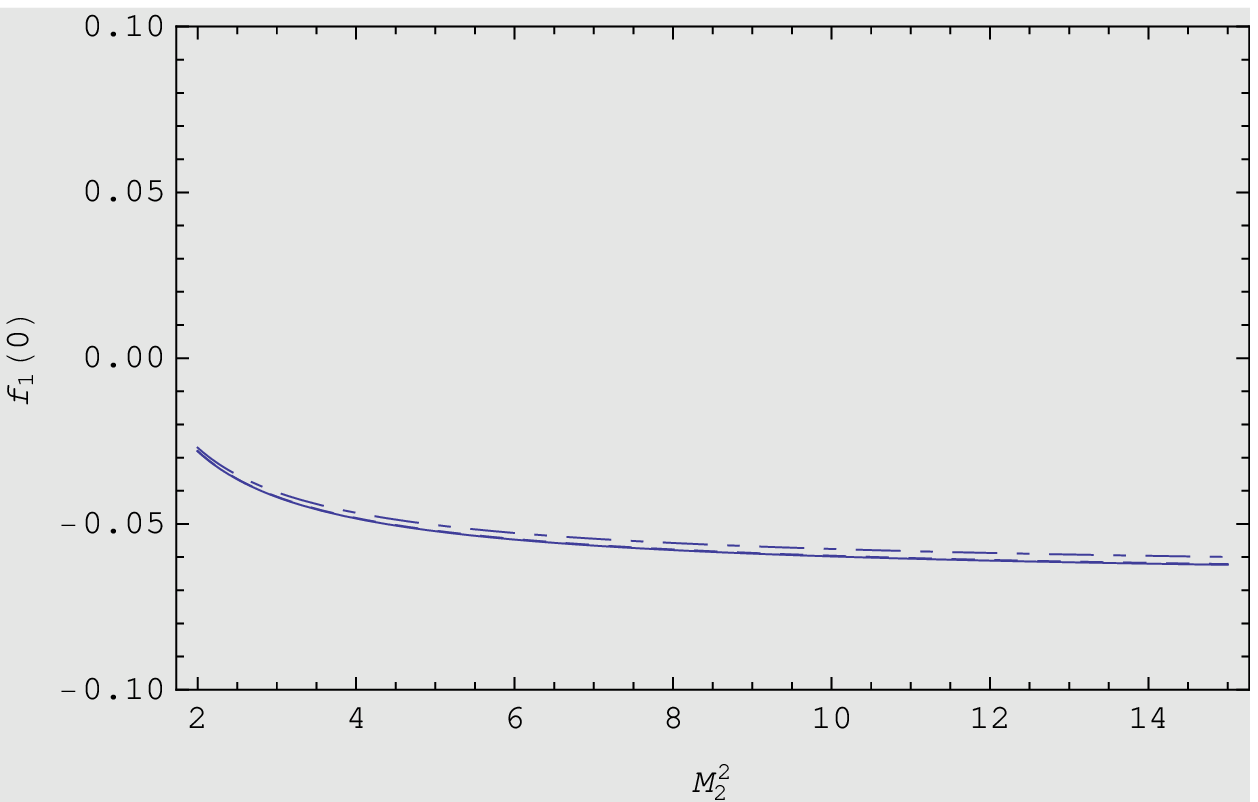}\hfill
\includegraphics[width=6cm]{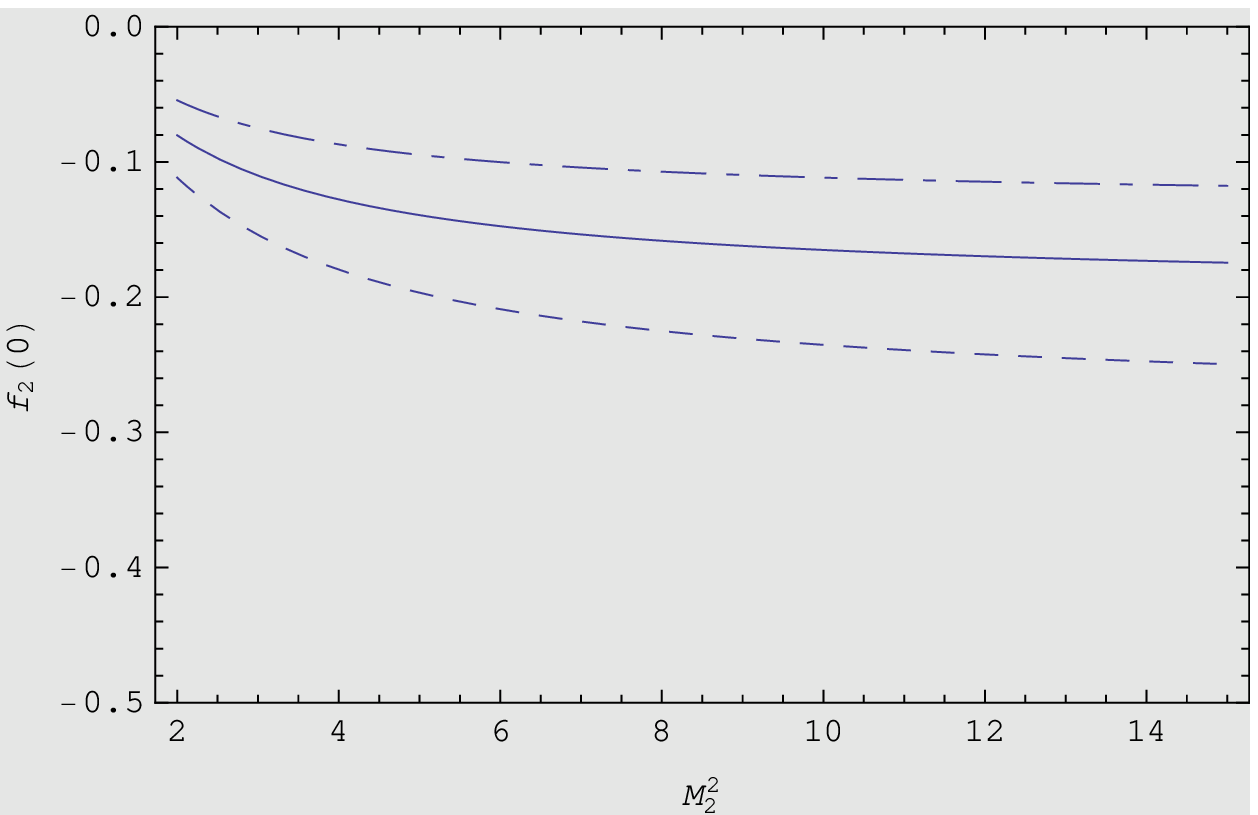}\\
\includegraphics[width=6cm]{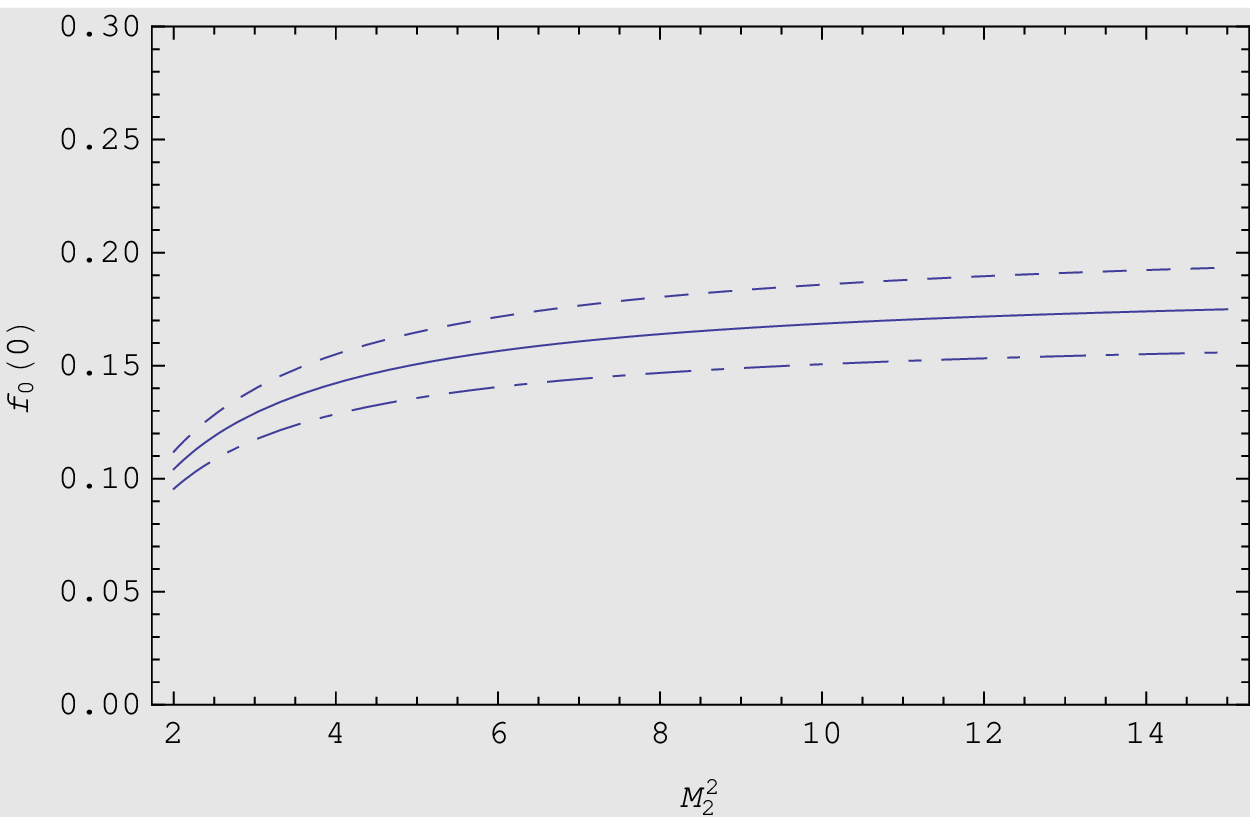}\hfill
\includegraphics[width=6cm]{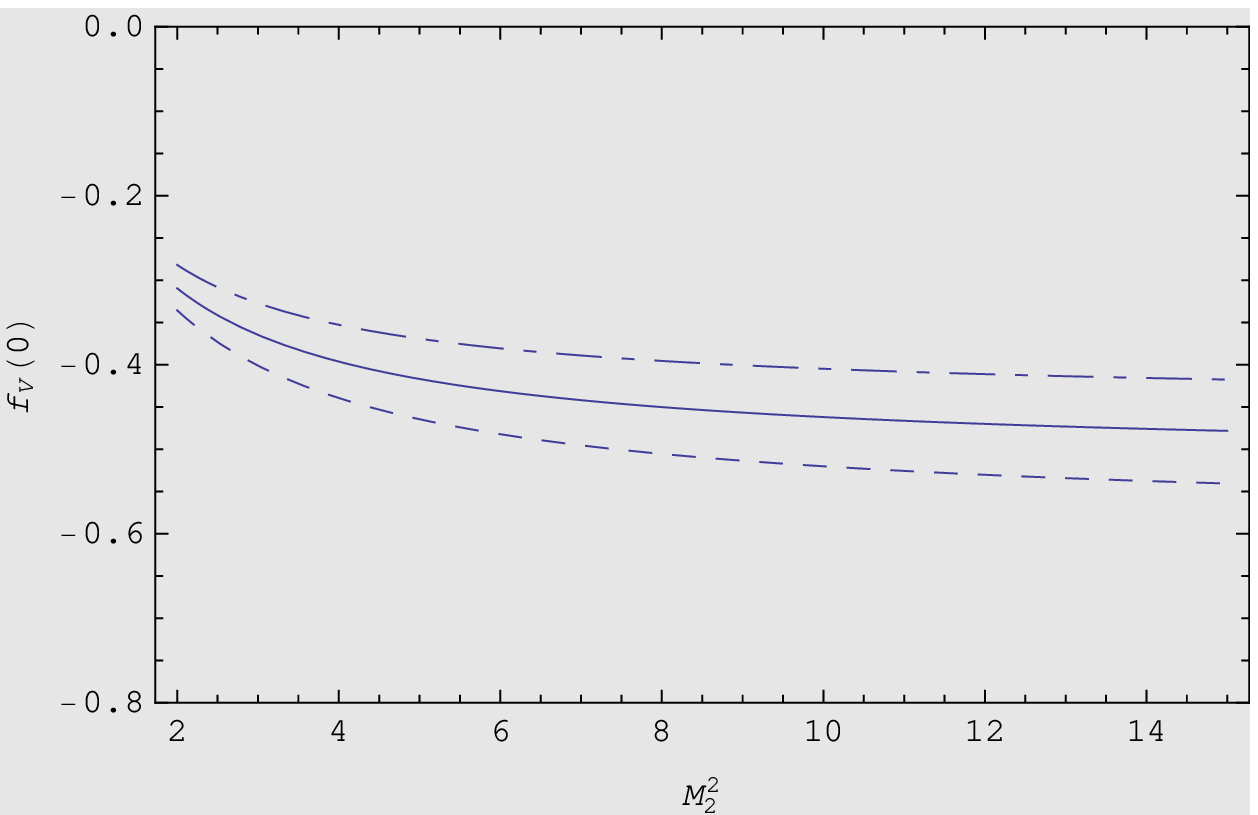}\\
\caption{ \label{FM2d0a1} $D^0 \rightarrow a_1^- (1260)$ transition
form factors at $q^2=0$ as functions of $M^2_2$. The dashed, solid
and dot dashed lines correspond to $s_0^\prime=$3.7, $3.5$ and $3.3$
${\rm GeV}^2$ respectively.}
\end{figure}
\begin{figure}[htbp]
\includegraphics[width=6cm]{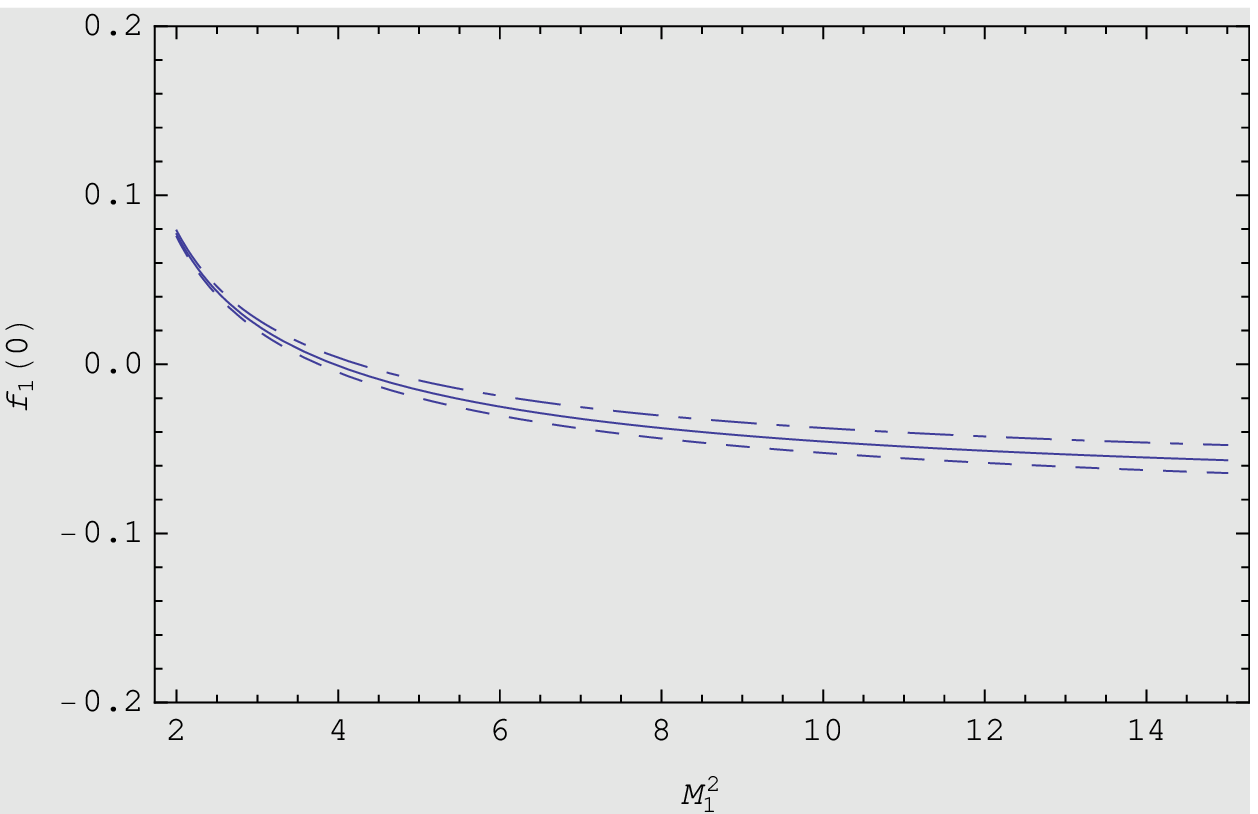}\hfill
\includegraphics[width=6cm]{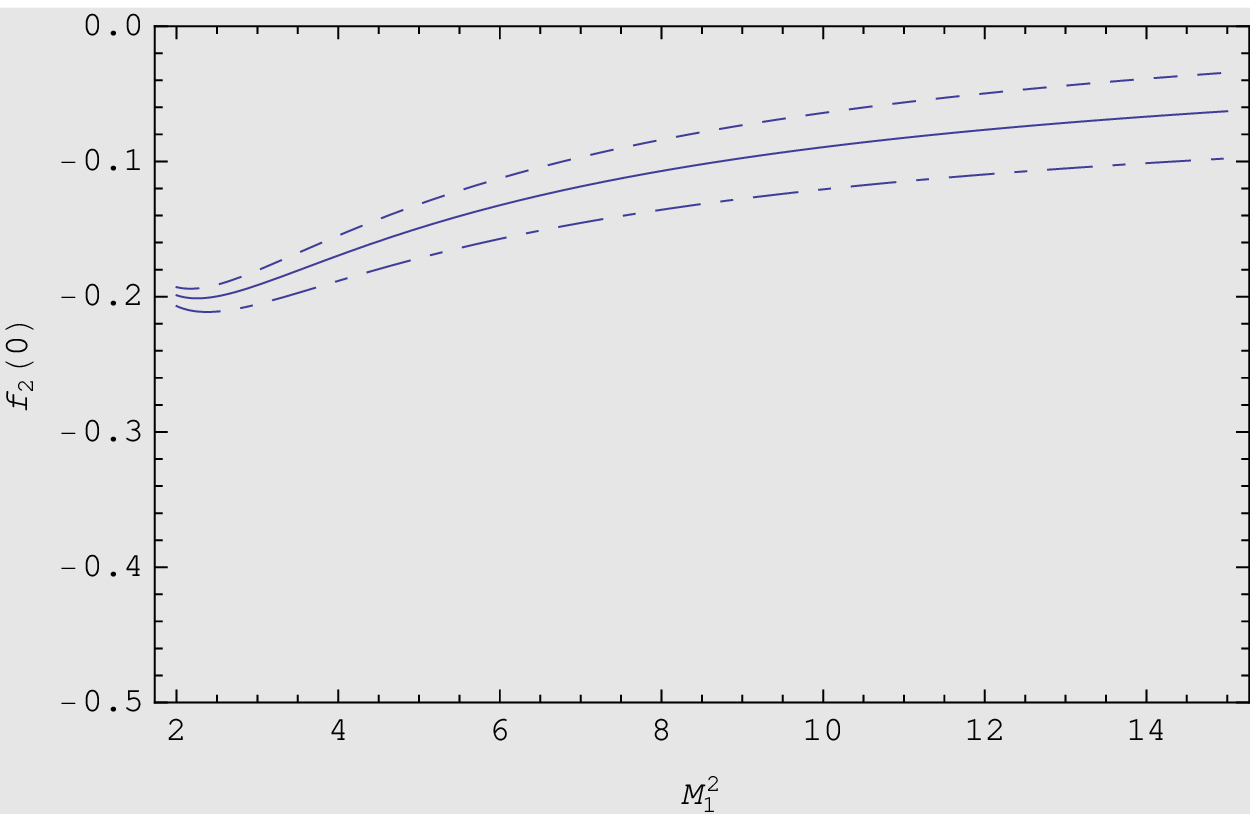}\\
\includegraphics[width=6cm]{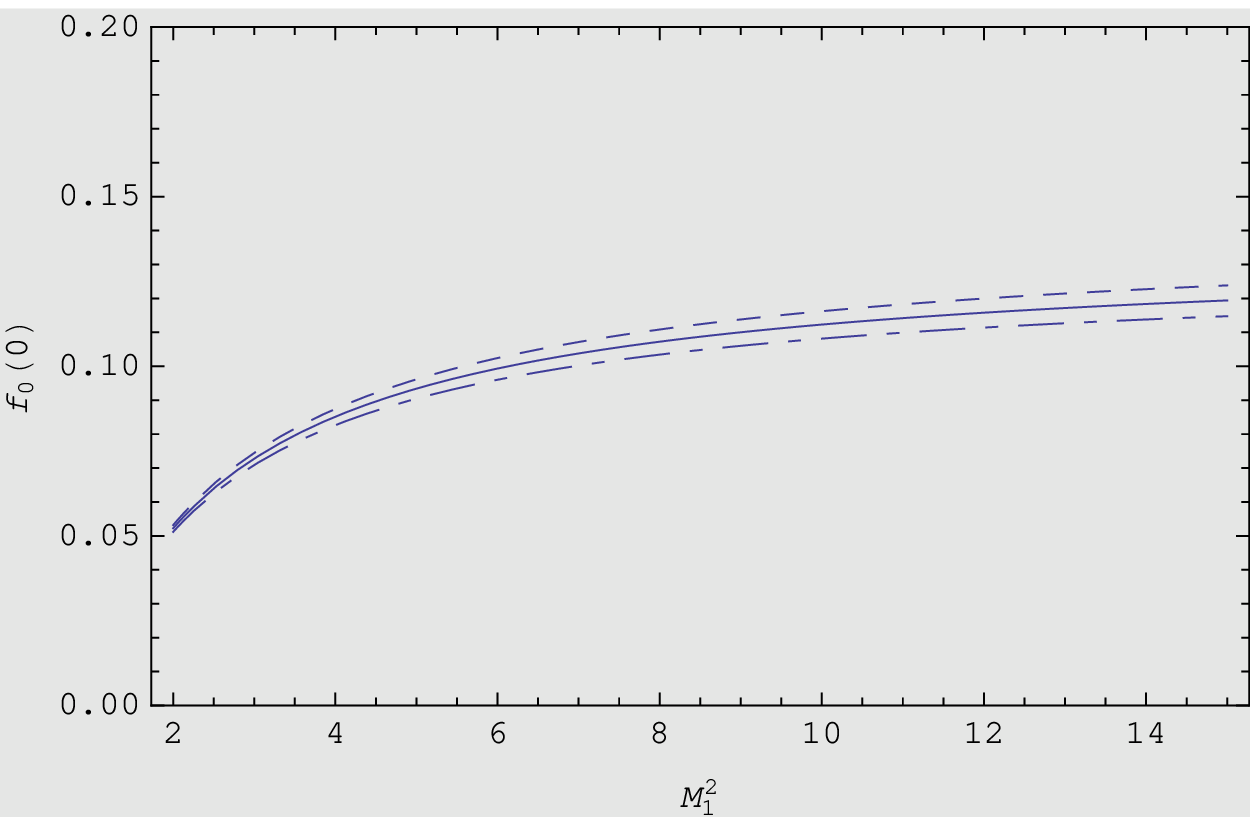}\hfill
\includegraphics[width=6cm]{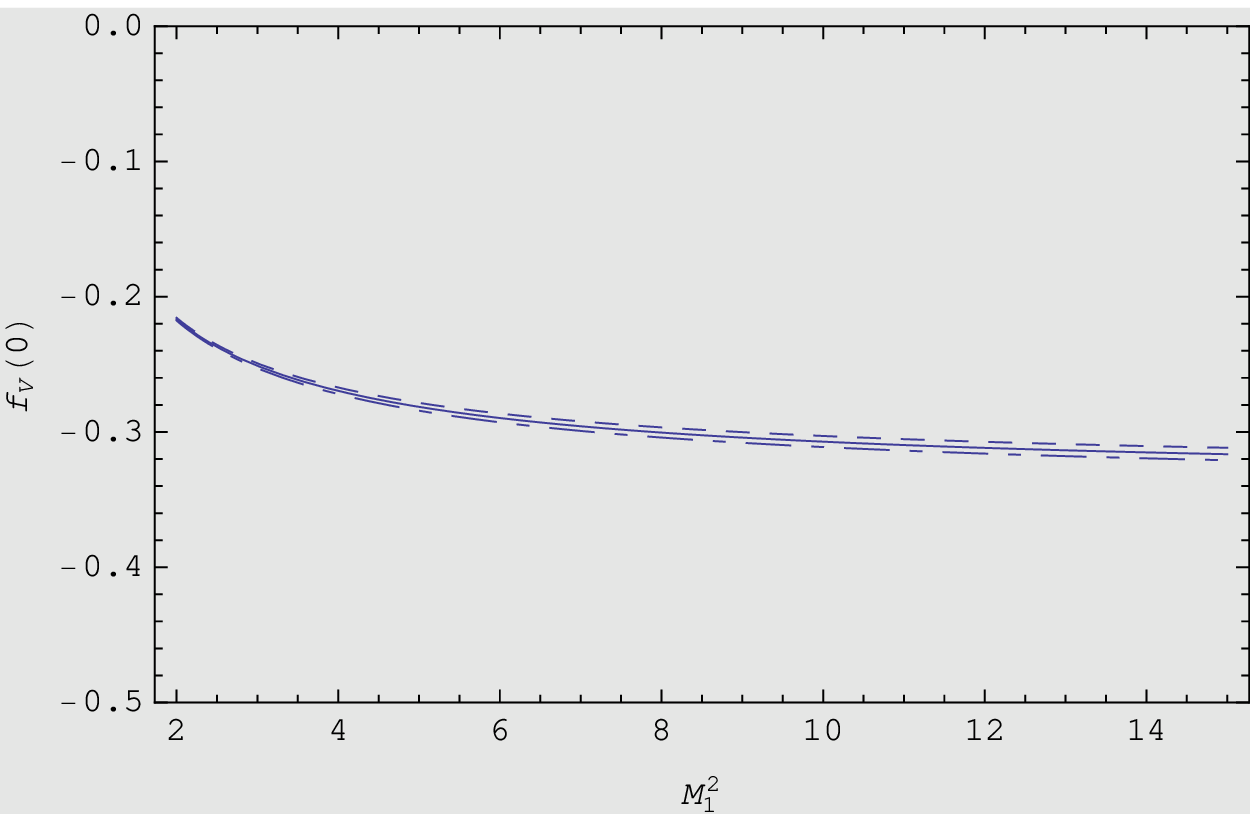}\\
\caption{ \label{FLM1dpf1} $D^+ \rightarrow f_1 (1285)$ transition
form factors at $q^2=0$ as functions of $M^2_1$. The dashed, solid
and dot dashed lines correspond to $s_0=7.2$, $7.0$ and $6.8$ ${\rm
GeV}^2$ respectively.}
\end{figure}
\begin{figure}[htbp]
\includegraphics[width=6cm]{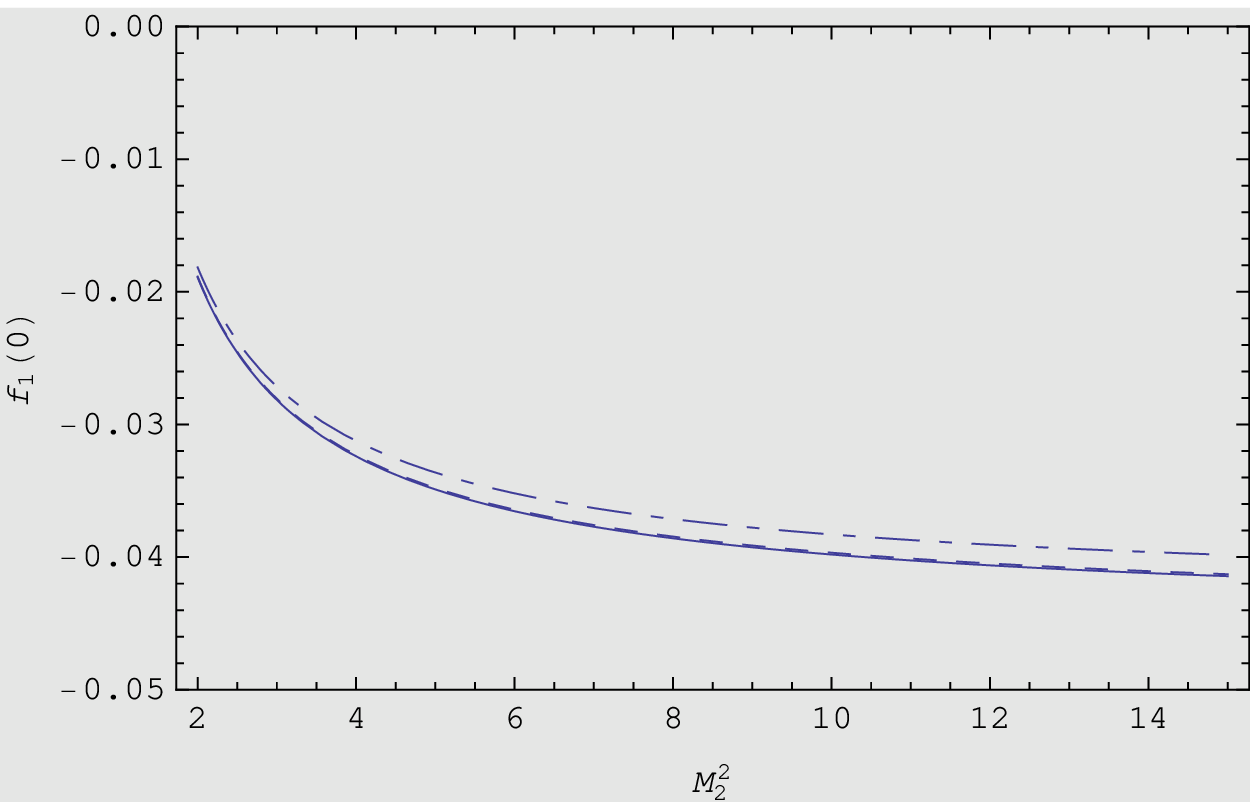}\hfill
\includegraphics[width=6cm]{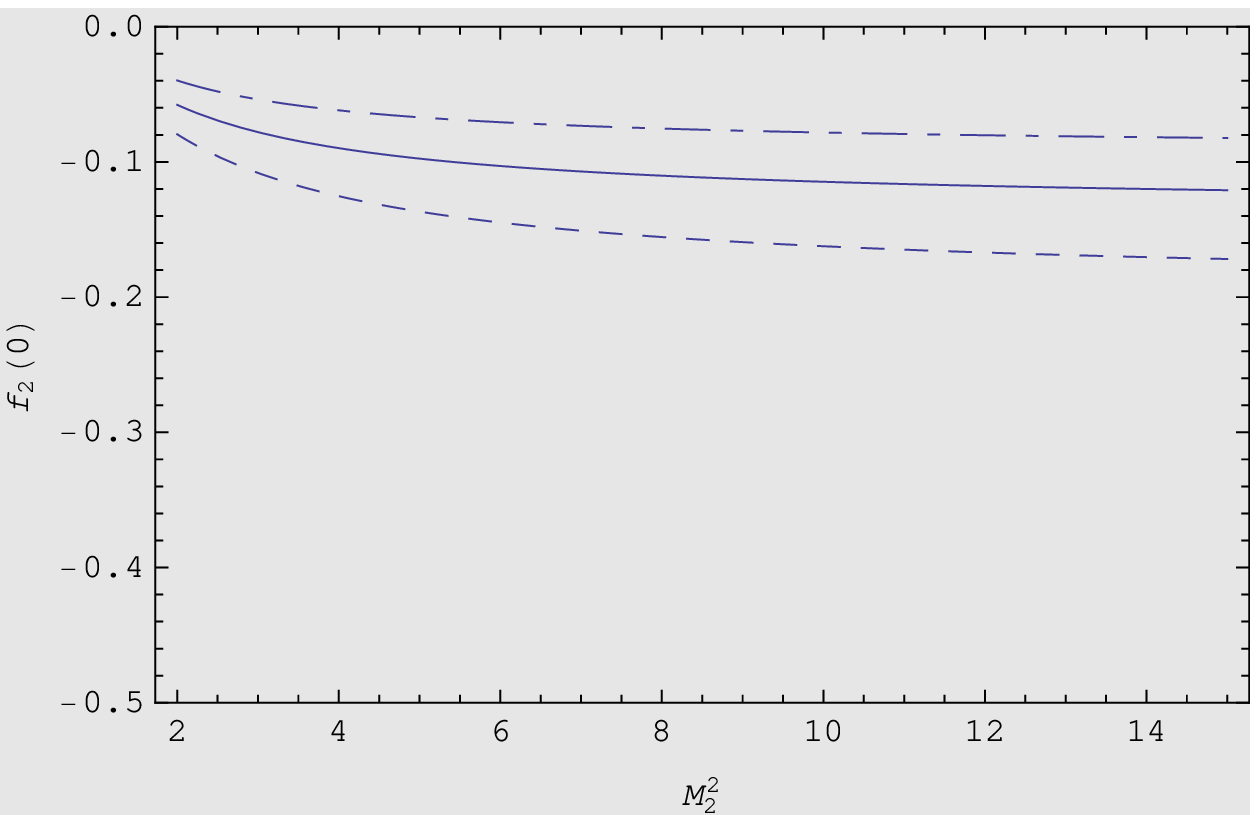}\\
\includegraphics[width=6cm]{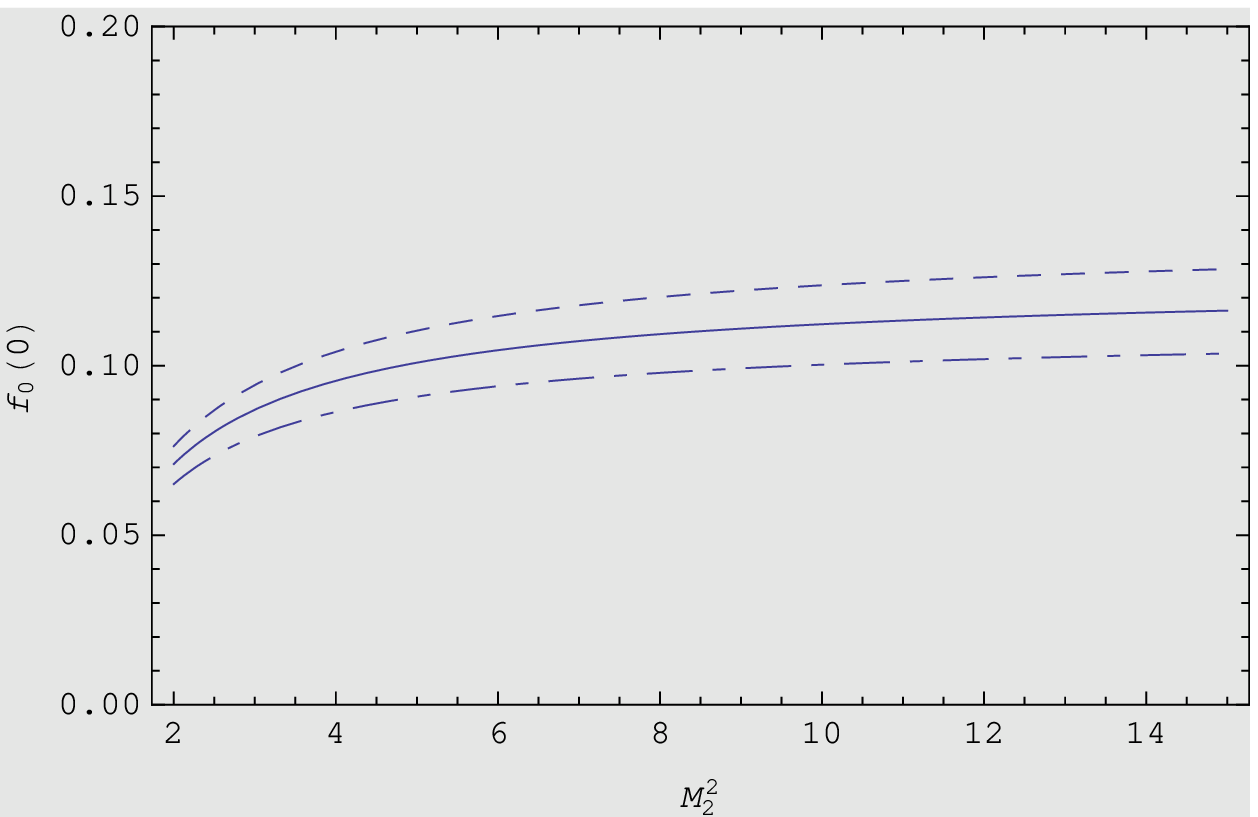}\hfill
\includegraphics[width=6cm]{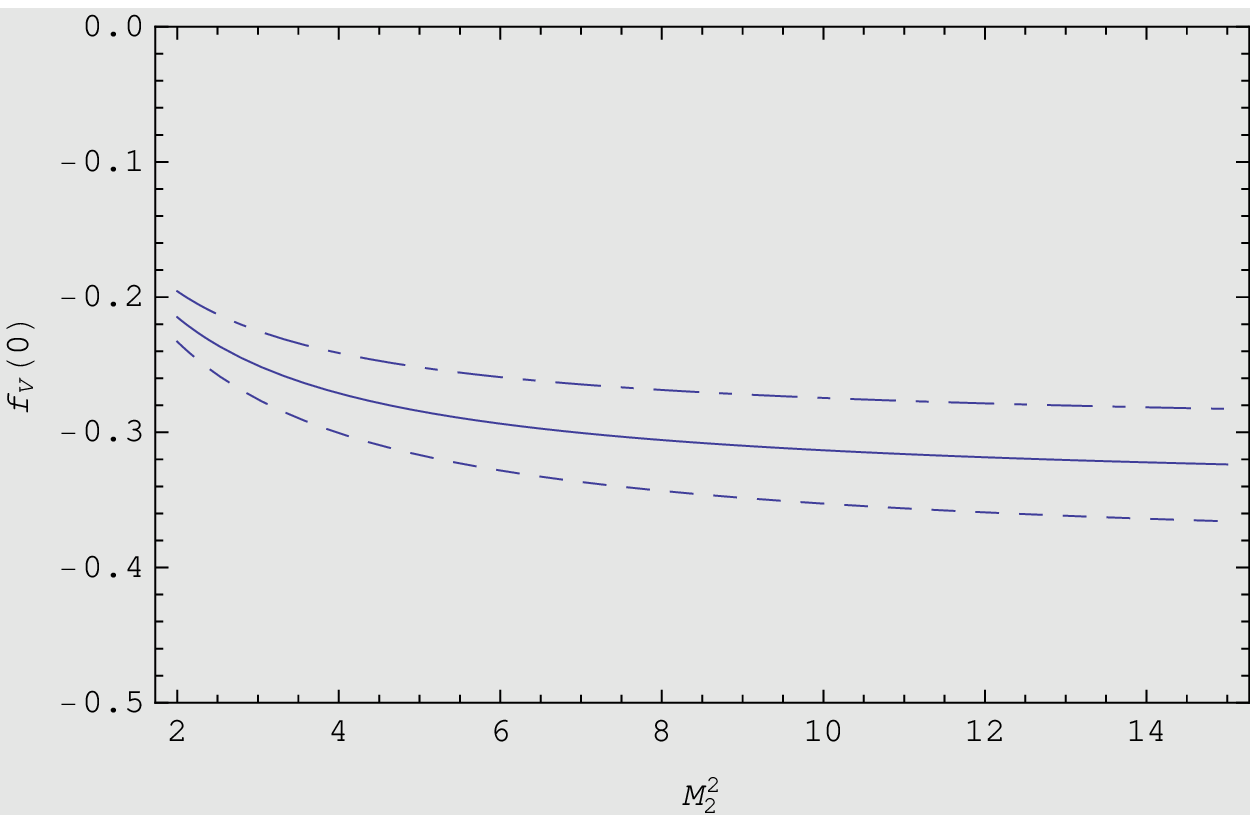}\\
\caption{ \label{FLM2dpf1} $D^+ \rightarrow f_1 (1285)$ transition
form factors at $q^2=0$ as functions of $M^2_2$. The dashed, solid
and dot dashed lines correspond to $s_0^\prime=3.7$, $3.5$ and $3.3$
${\rm GeV}^2$ respectively.}
\end{figure}
\begin{figure}[htbp]
\includegraphics[width=6cm]{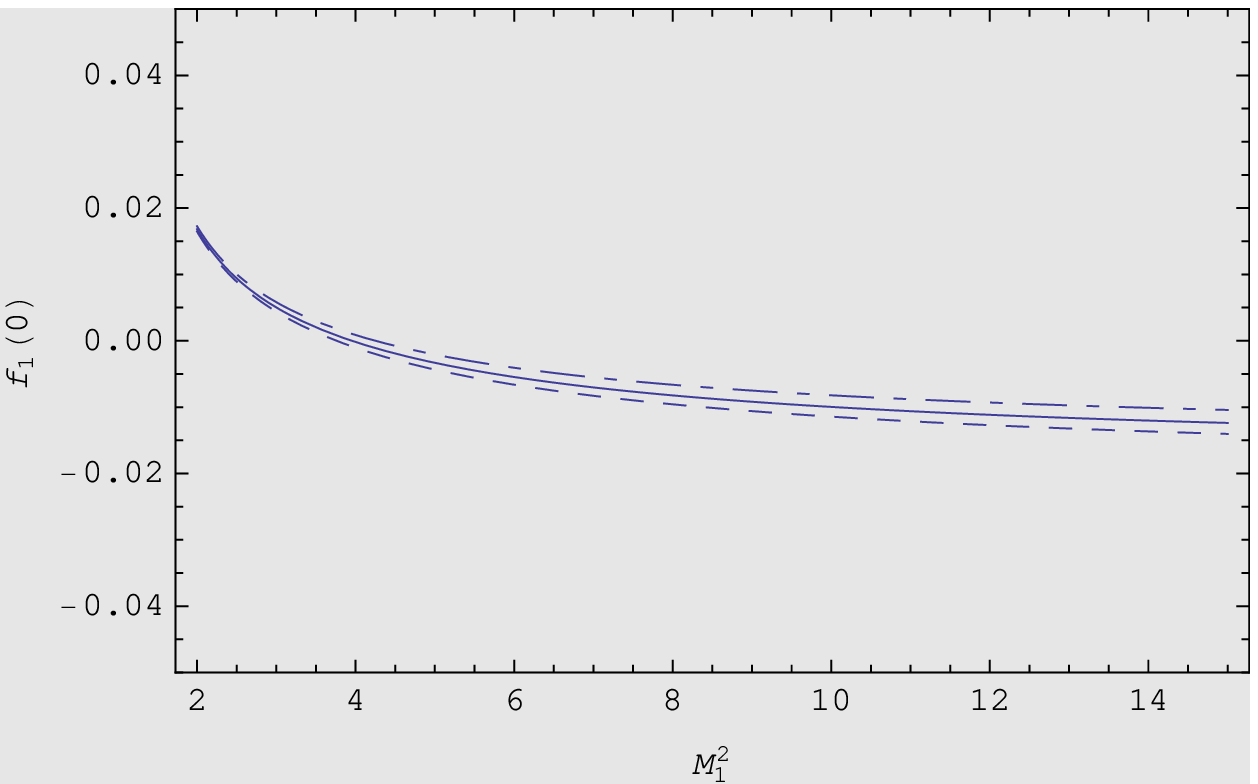}\hfill
\includegraphics[width=6cm]{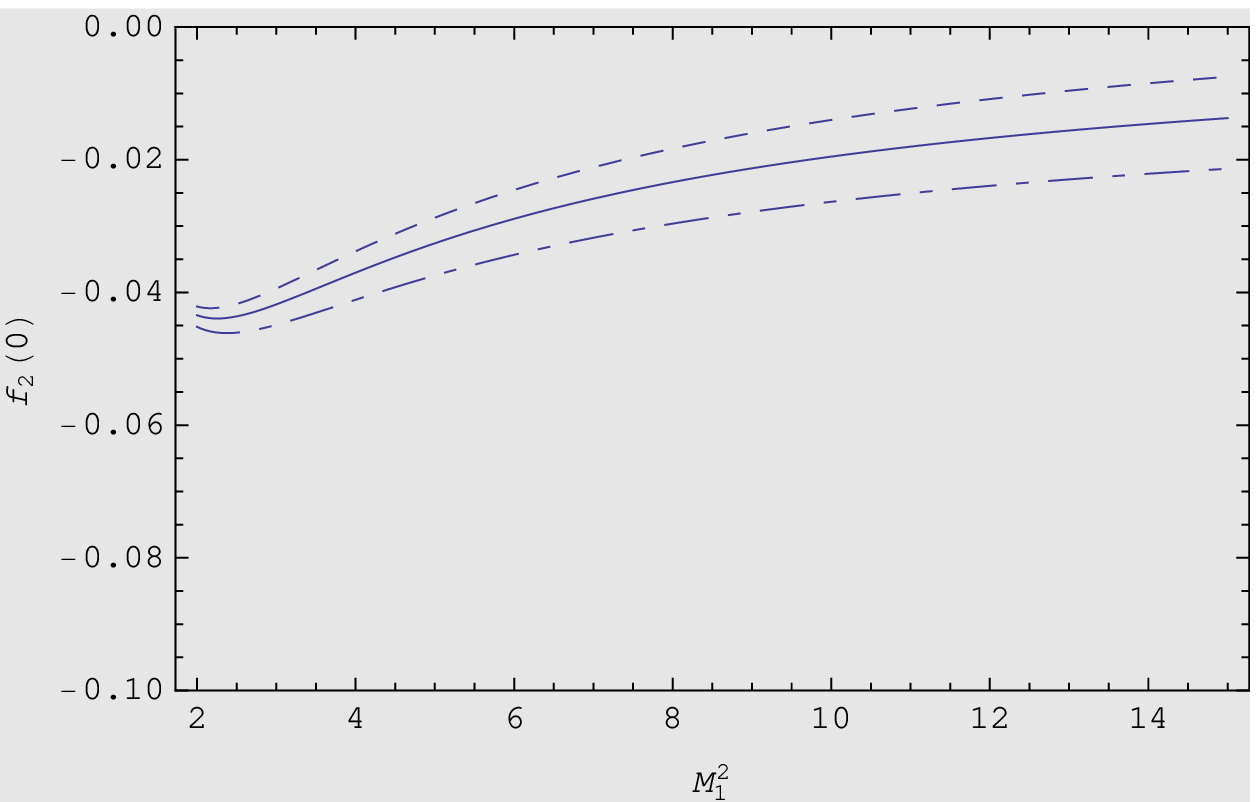}\\
\includegraphics[width=6cm]{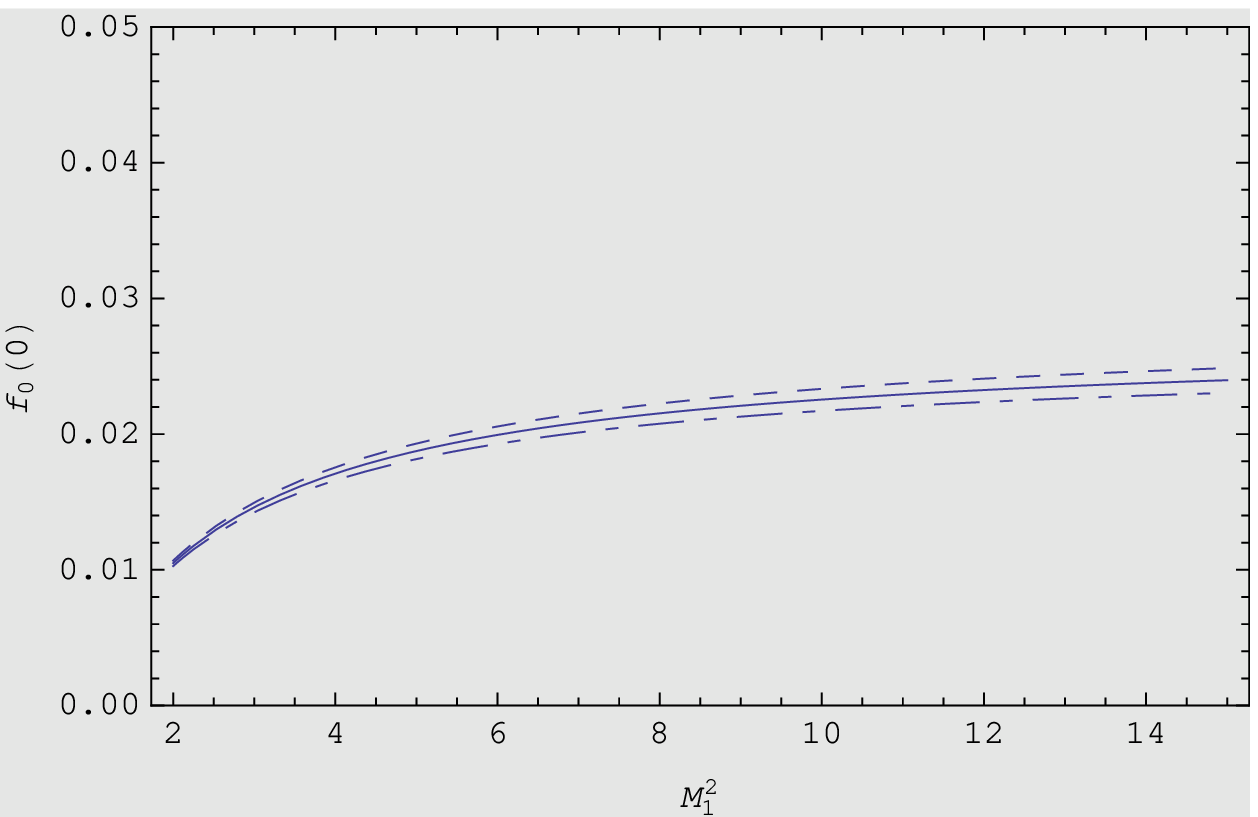}\hfill
\includegraphics[width=6cm]{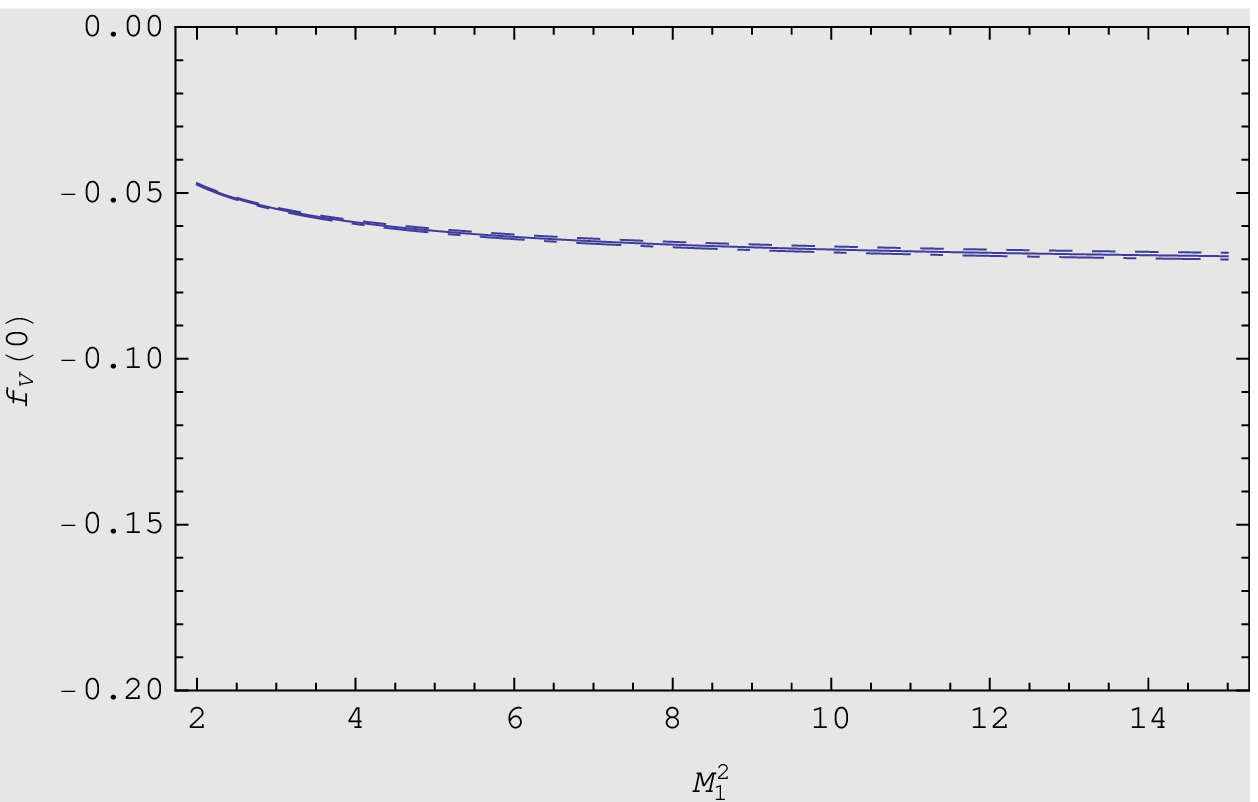}\\
\caption{ \label{FHM1dpf1} $D^+ \rightarrow f_1 (1420)$ transition
form factors at $q^2=0$ as functions of $M^2_1$. The dashed, solid
and dot dashed lines correspond to $s_0=7.2$, $7.0$ and $6.8$ ${\rm
GeV}^2$ respectively.}
\end{figure}
\begin{figure}[htbp]
\includegraphics[width=6cm]{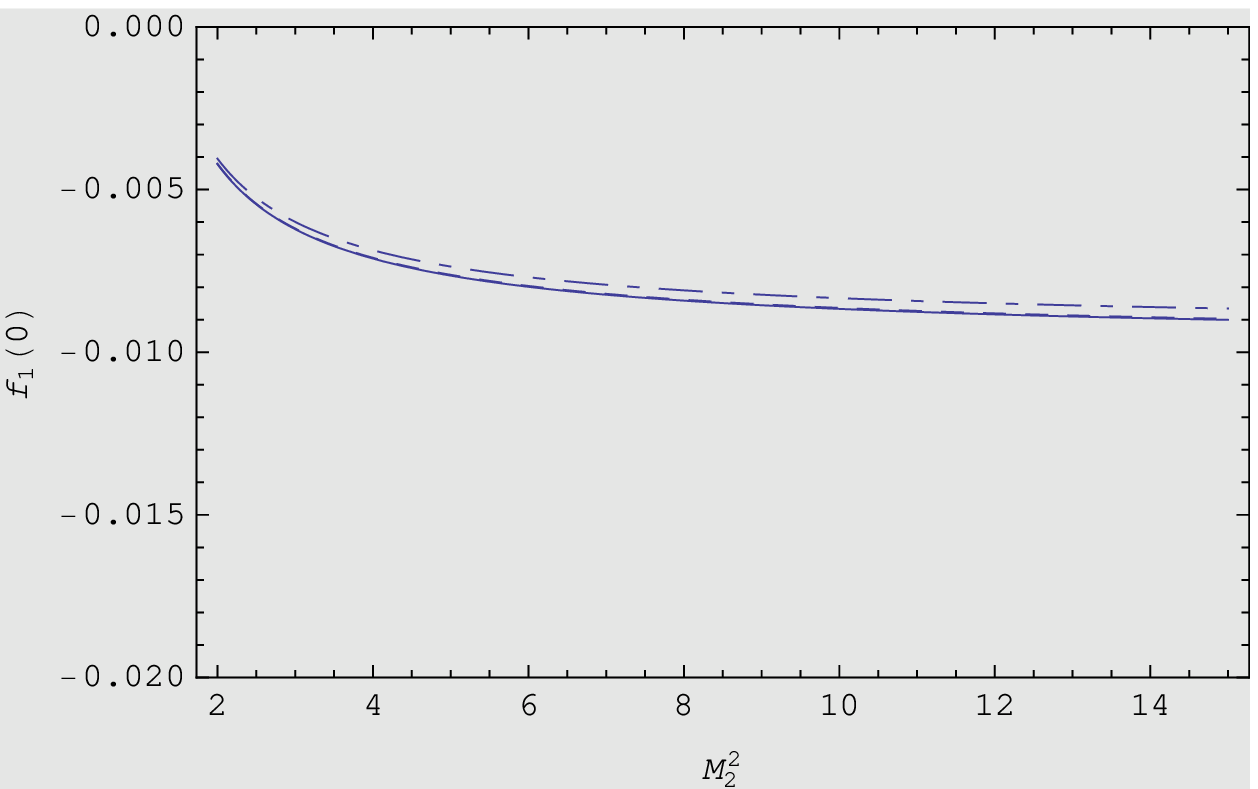}\hfill
\includegraphics[width=6cm]{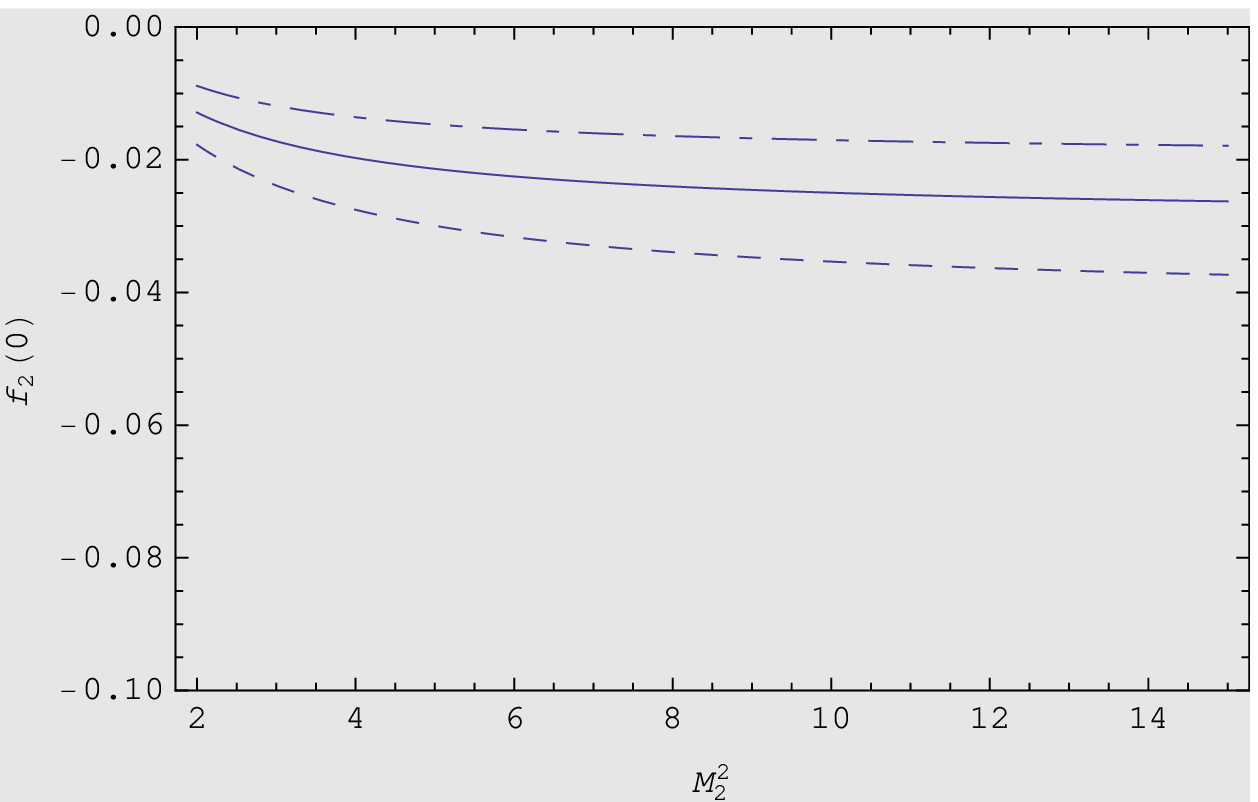}\\
\includegraphics[width=6cm]{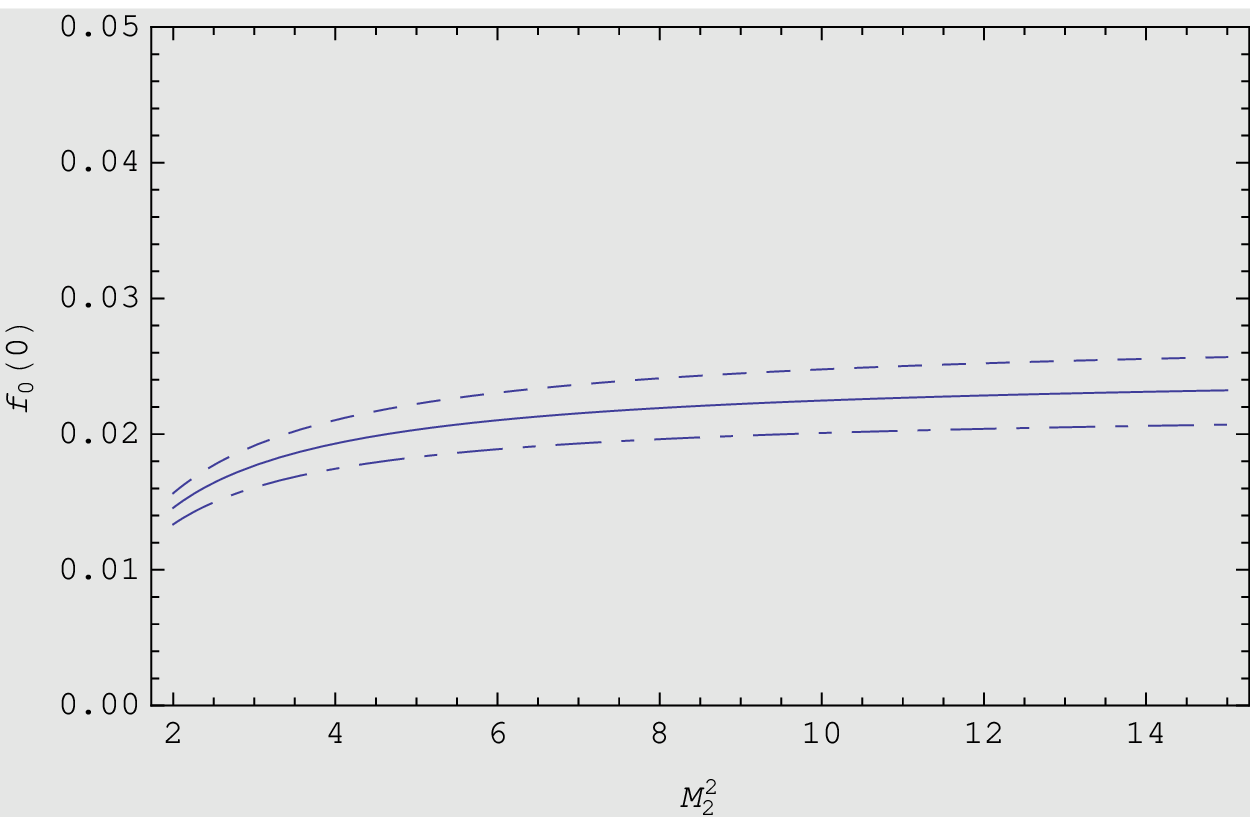}\hfill
\includegraphics[width=6cm]{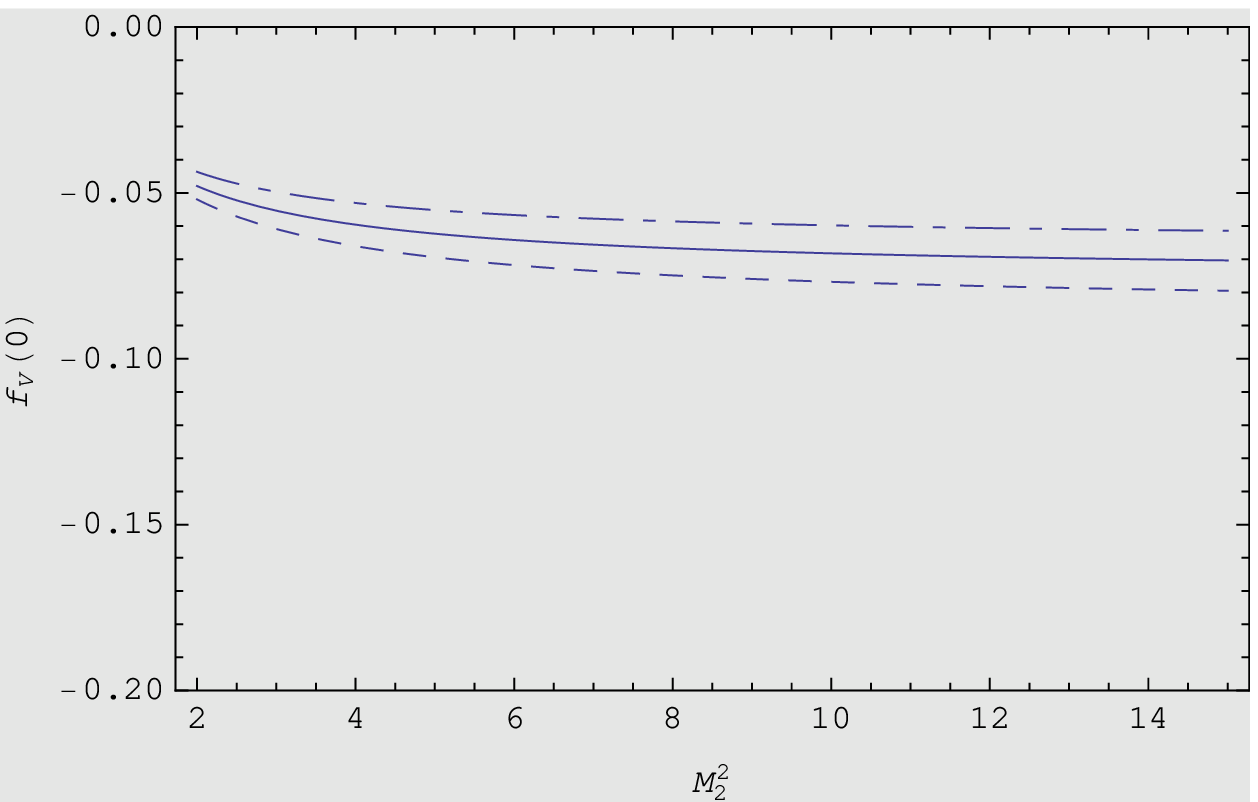}\\
\caption{ \label{FHM2dpf1} $D^+ \rightarrow f_1 (1420)$ transition
form factors at $q^2=0$ as functions of $M^2_2$. The dashed, solid
and dot dashed lines correspond to $s_0^\prime=3.7$, $3.5$ and $3.3$
${\rm GeV}^2$ respectively.}
\end{figure}

 From Eq.(\ref{sumrule}), it is easily seen that the
sum rules for form factors contain four free parameters $M^2_1$,
$M^2_2$, $s_0$, $s^\prime_0$. $M^2_1$ and $M^2_2$ are Borel mass
squares. $s_0$ and $s^\prime_0$ are the continuum thresholds of $D$
and $A$ mesons respectively. These are not physical quantities, so
the form factors as physical quantities should be independent of
them. Practically, the allowed regions for $M^2_1$, $M^2_2$, $s_0$,
$s^\prime_0$ are determined by requiring the curves of form factors
as functions of these four parameters to be most stable. As
illustrations, we give the $D^+ \rightarrow a_1^0 (1260)$ transition
form factors at zero momentum transfer ($q^2=0$) as functions of
Borel mass squares $M^2_1$ in Fig.\ref{FM1dpa1}. The corresponding
variation of form factors with $M^2_2$ and those for $D^0
\rightarrow a_1^- (1260), D^+ \rightarrow f_1 (1285), f_1( 1420)$
decays are similar and shown in Fig.\ref{FM2dpa1}-\ref{FHM2dpf1}.

\begin{figure}[htbp]
\includegraphics[width=6cm]{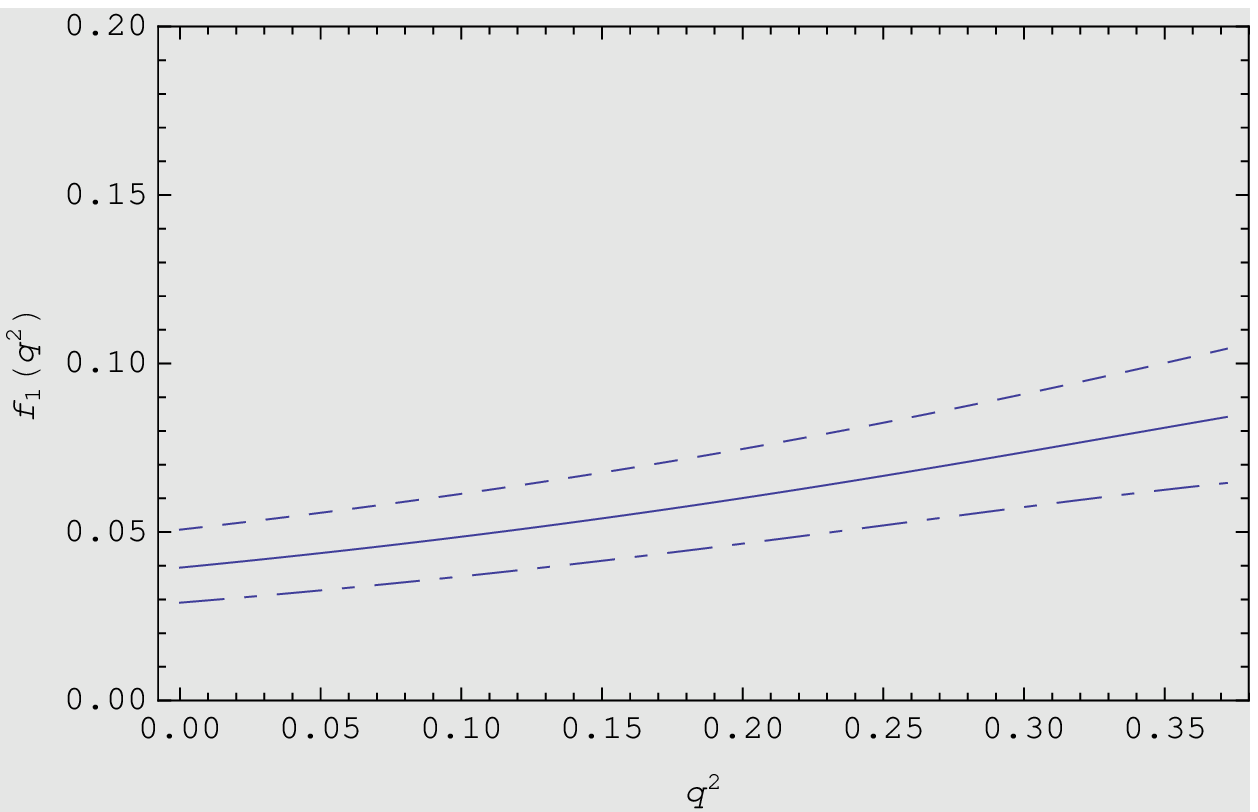}\hfill
\includegraphics[width=6cm]{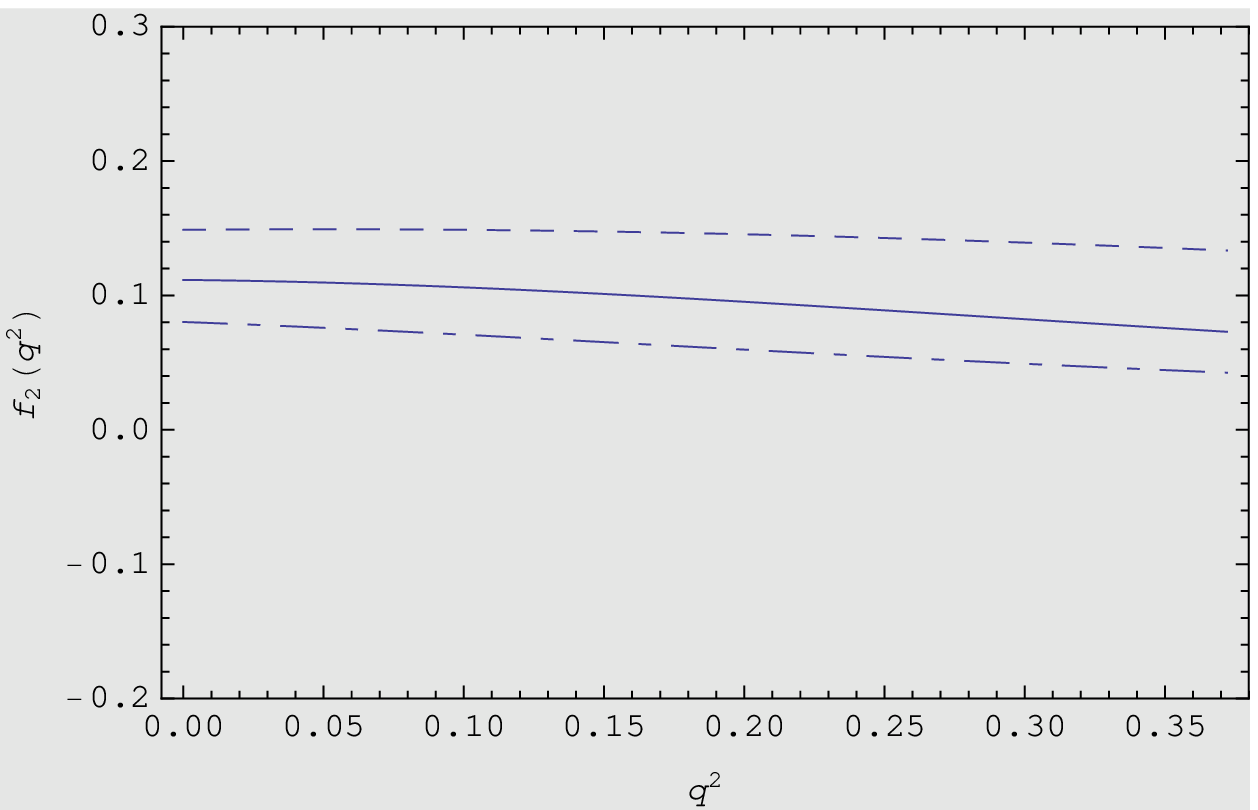}\\
\includegraphics[width=6cm]{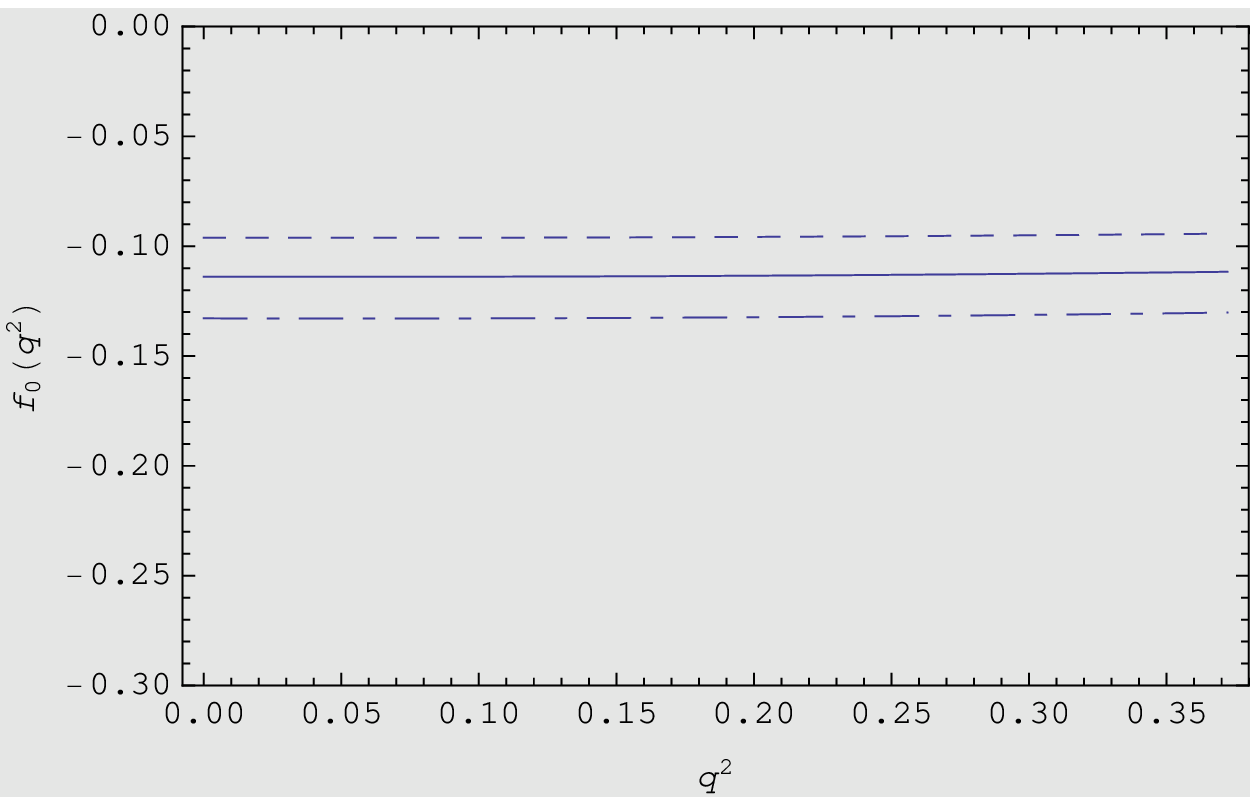}\hfill
\includegraphics[width=6cm]{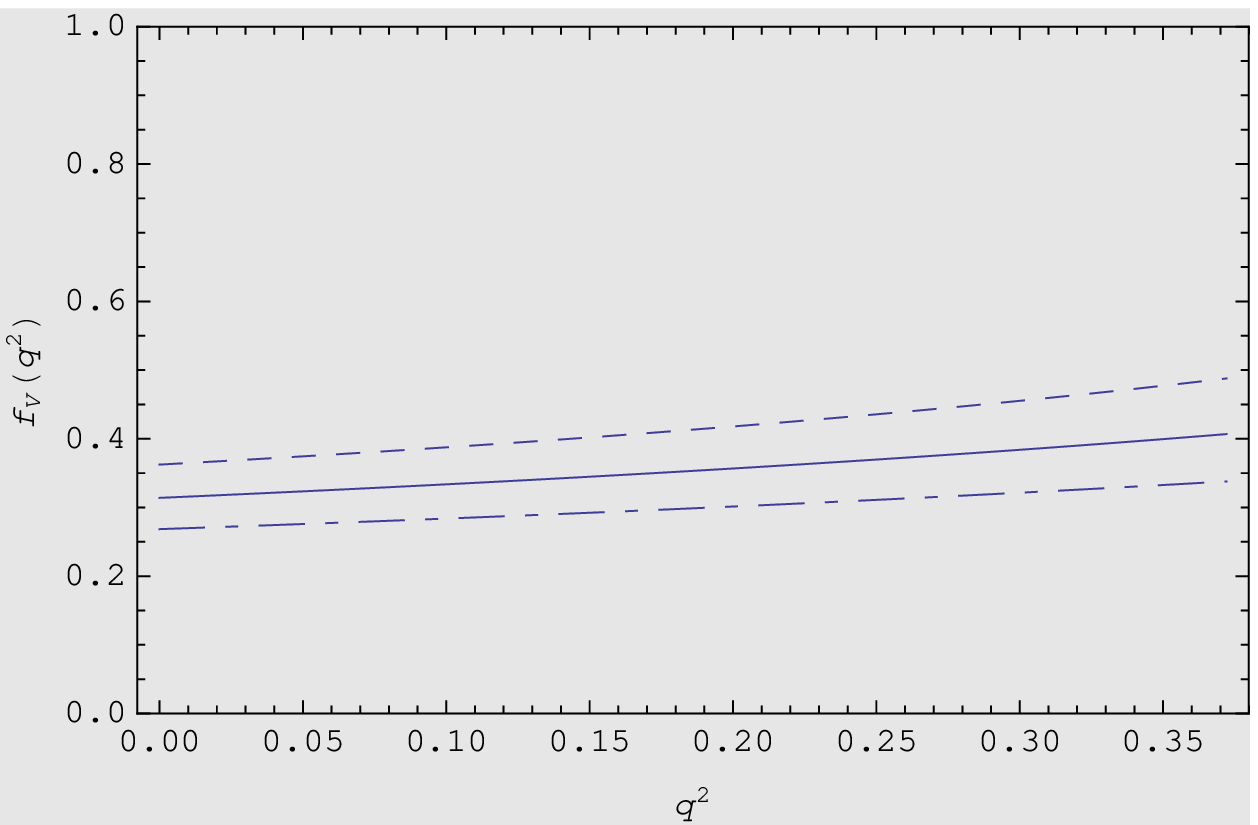}\\
\caption{ \label{Fqdpa1} The variation of $D^+ \rightarrow a_1^0
(1260)$ transition form factors as functions of $q^2$. The solid,
dot dashed and dashed lines correspond to the center value, upper
and lower limit respectively.}
\end{figure}

\begin{figure}[htbp]
\includegraphics[width=6cm]{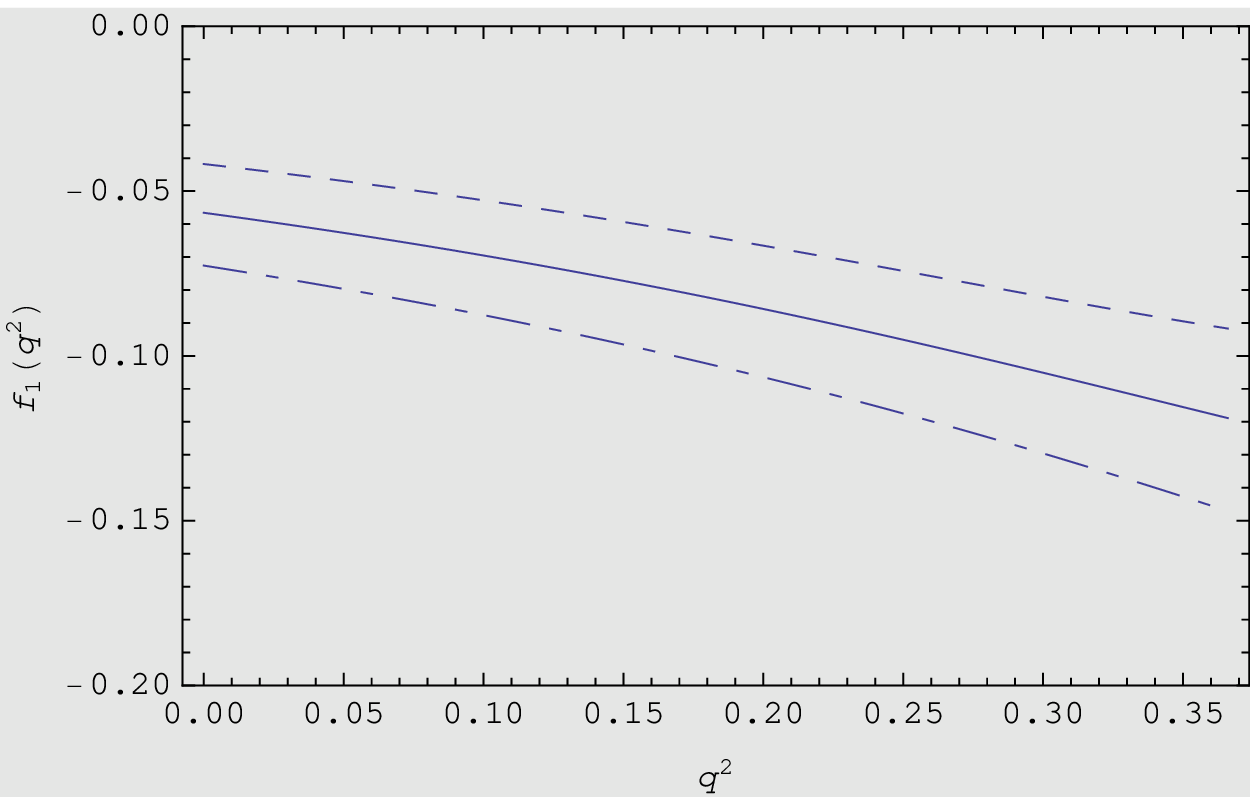}\hfill
\includegraphics[width=6cm]{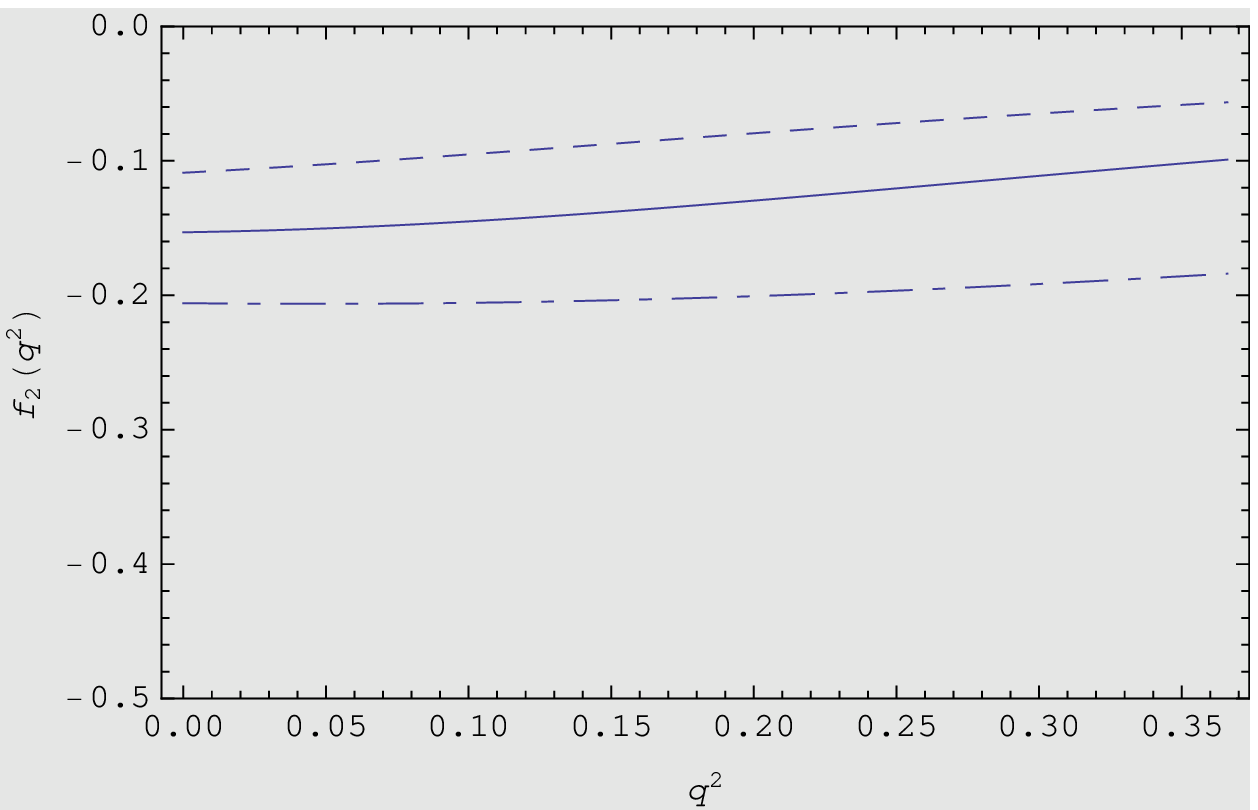}\\
\includegraphics[width=6cm]{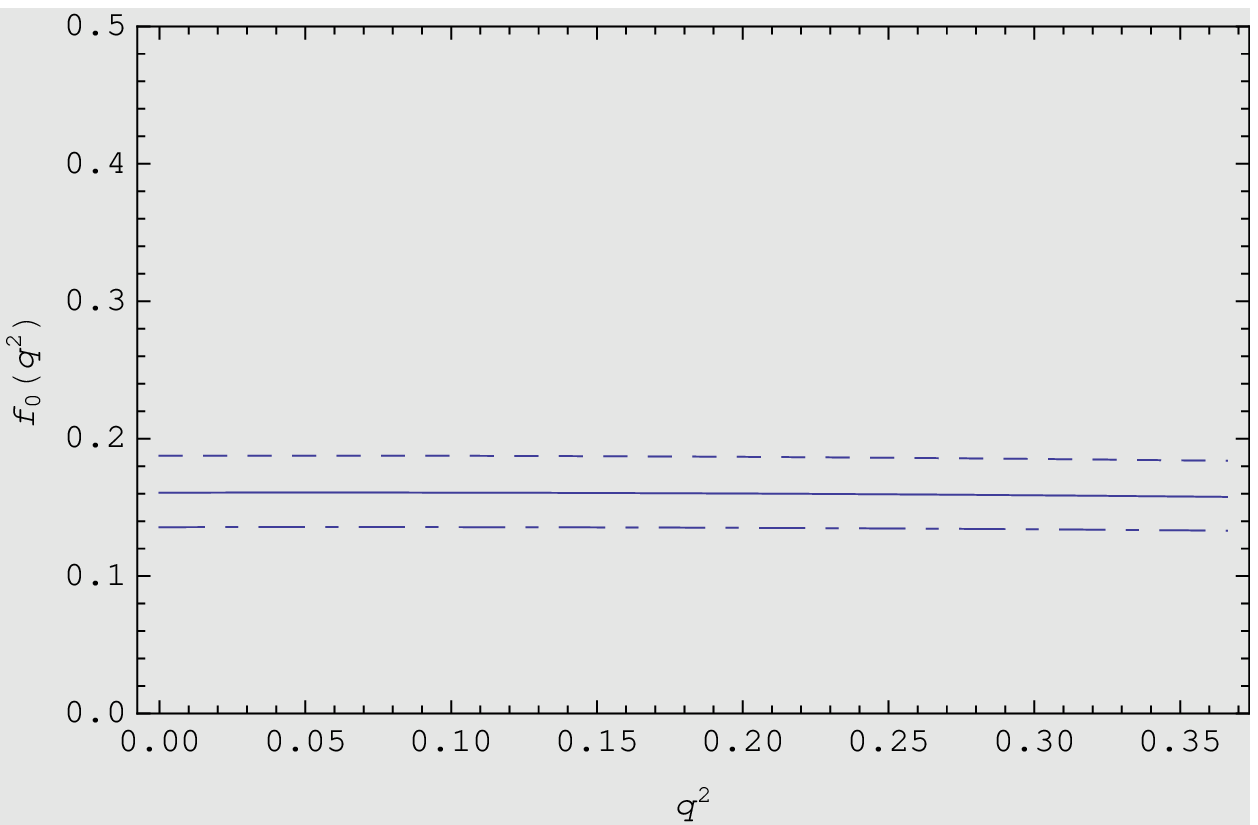}\hfill
\includegraphics[width=6cm]{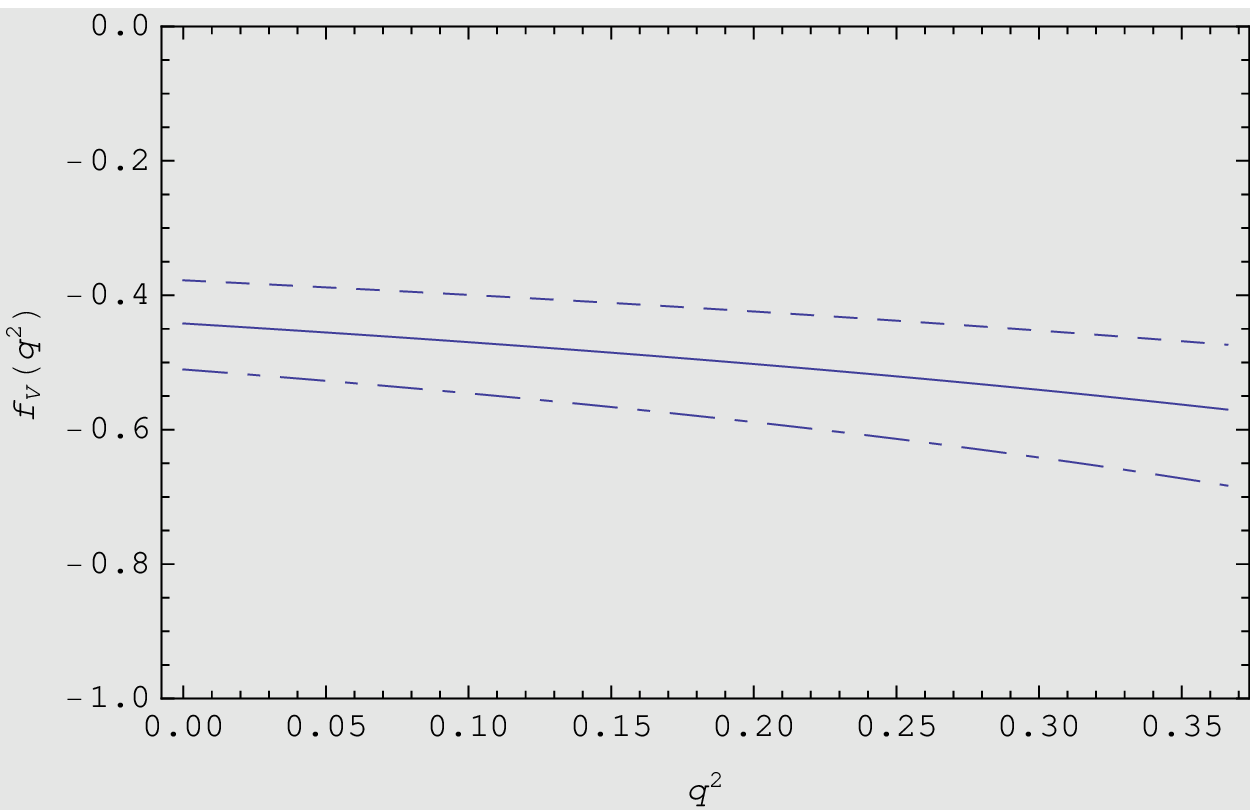}\\
\caption{ \label{Fqd0a1} The variation of $D^0 \rightarrow a_1^-
(1260)$ transition form factors as functions of $q^2$. The solid,
dot dashed and dashed lines correspond to the center value, upper
and lower limit respectively.}
\end{figure}

\begin{figure}[htbp]
\includegraphics[width=6cm]{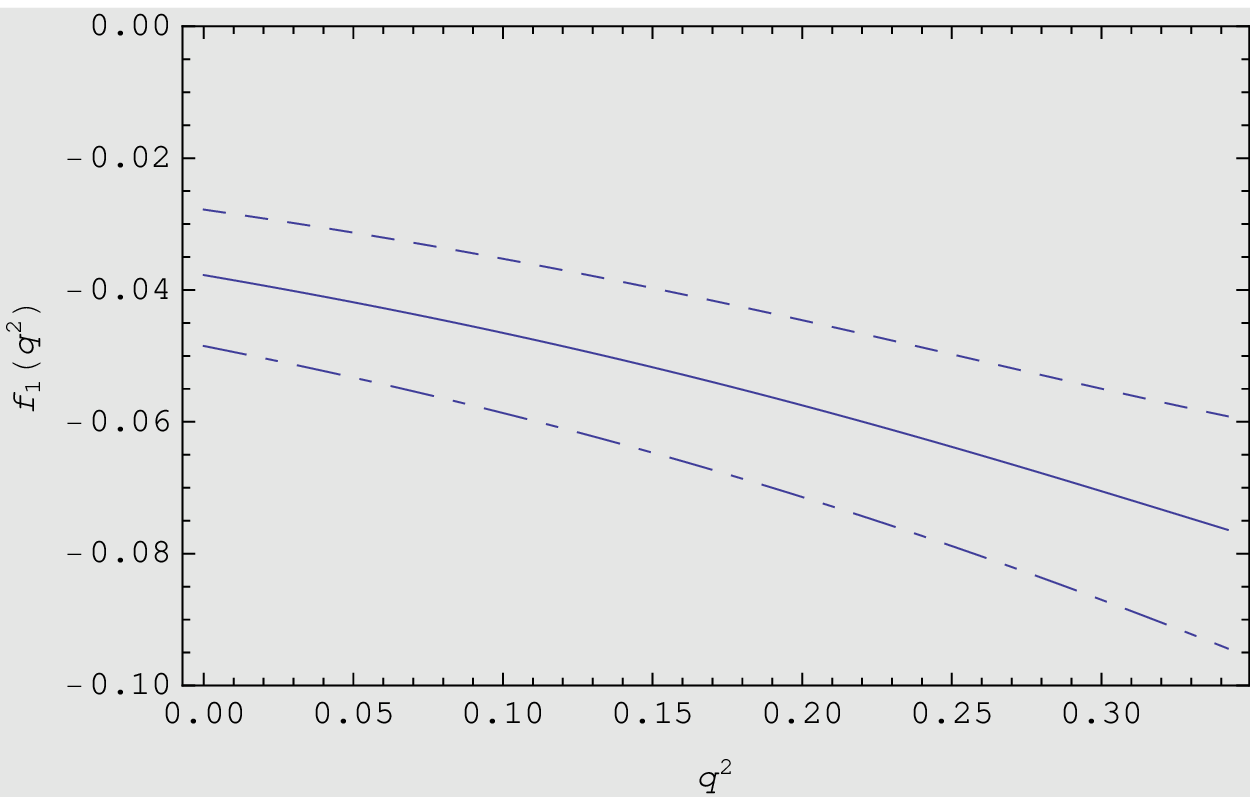}\hfill
\includegraphics[width=6cm]{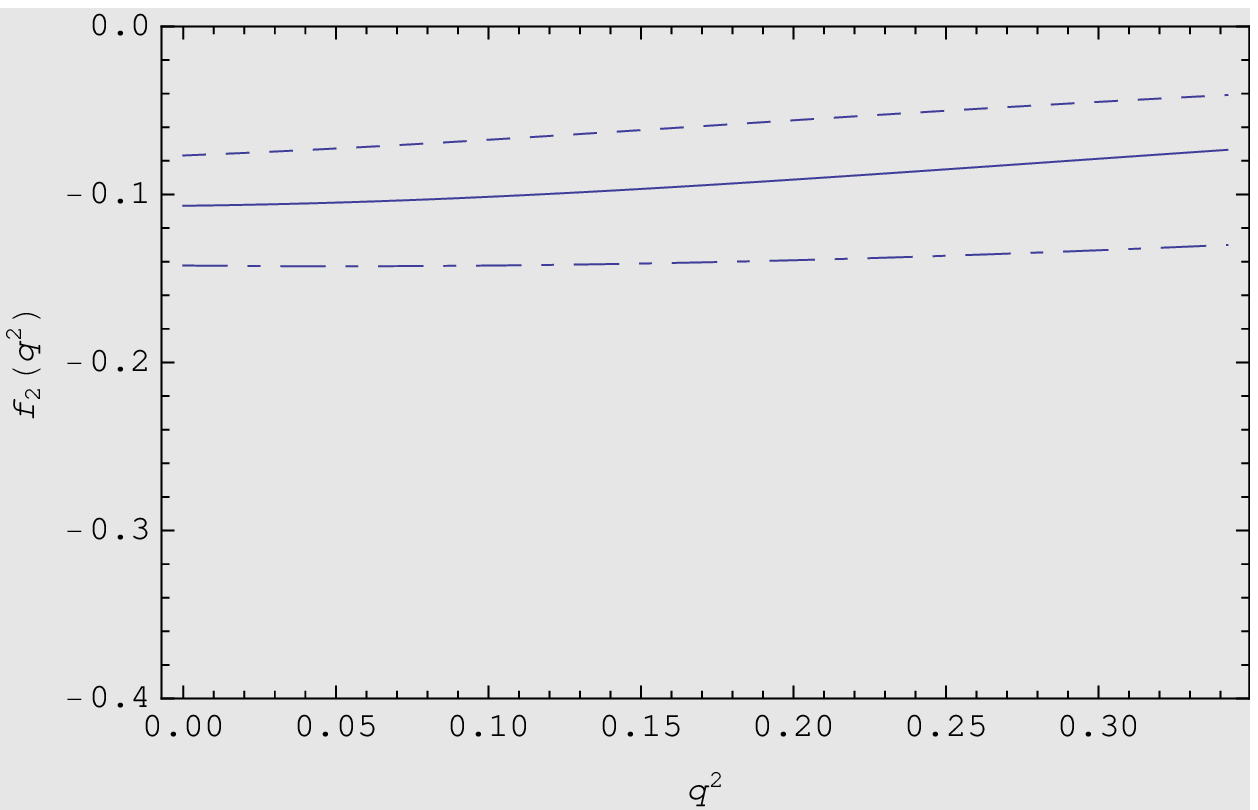}\\
\includegraphics[width=6cm]{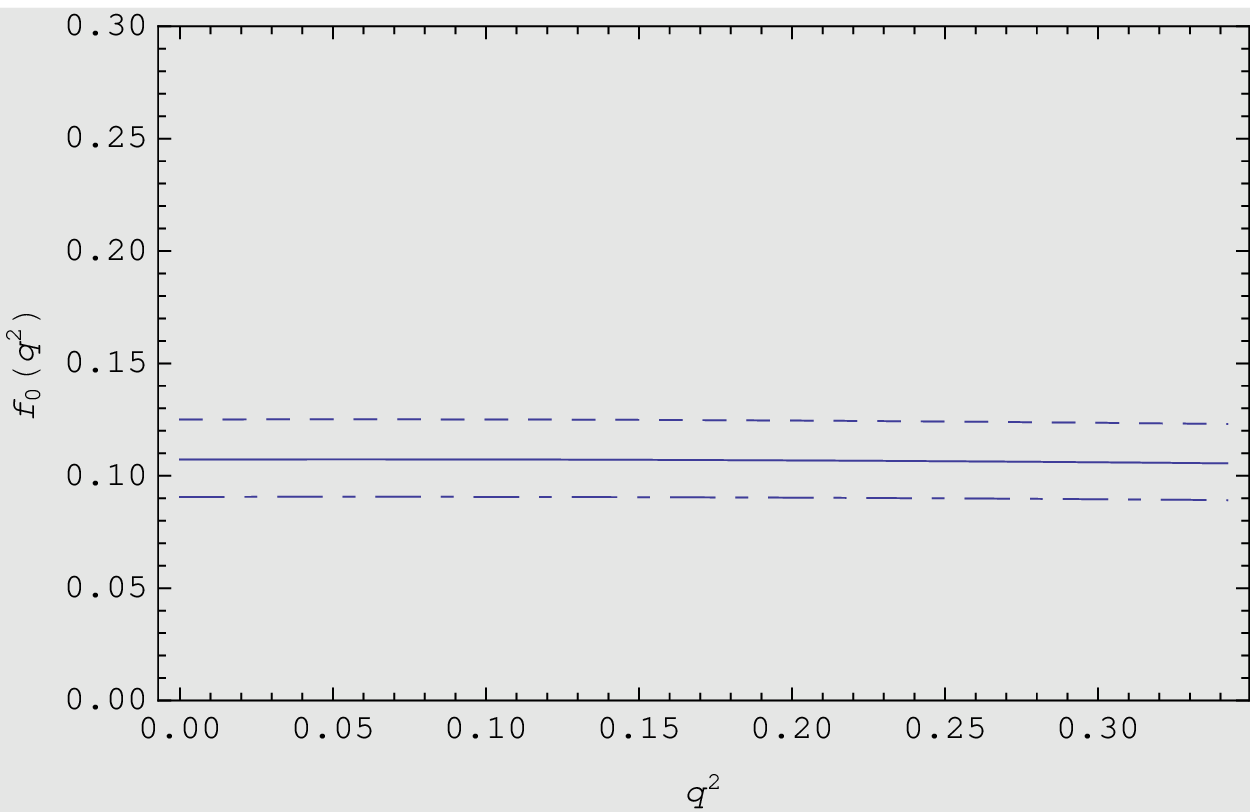}\hfill
\includegraphics[width=6cm]{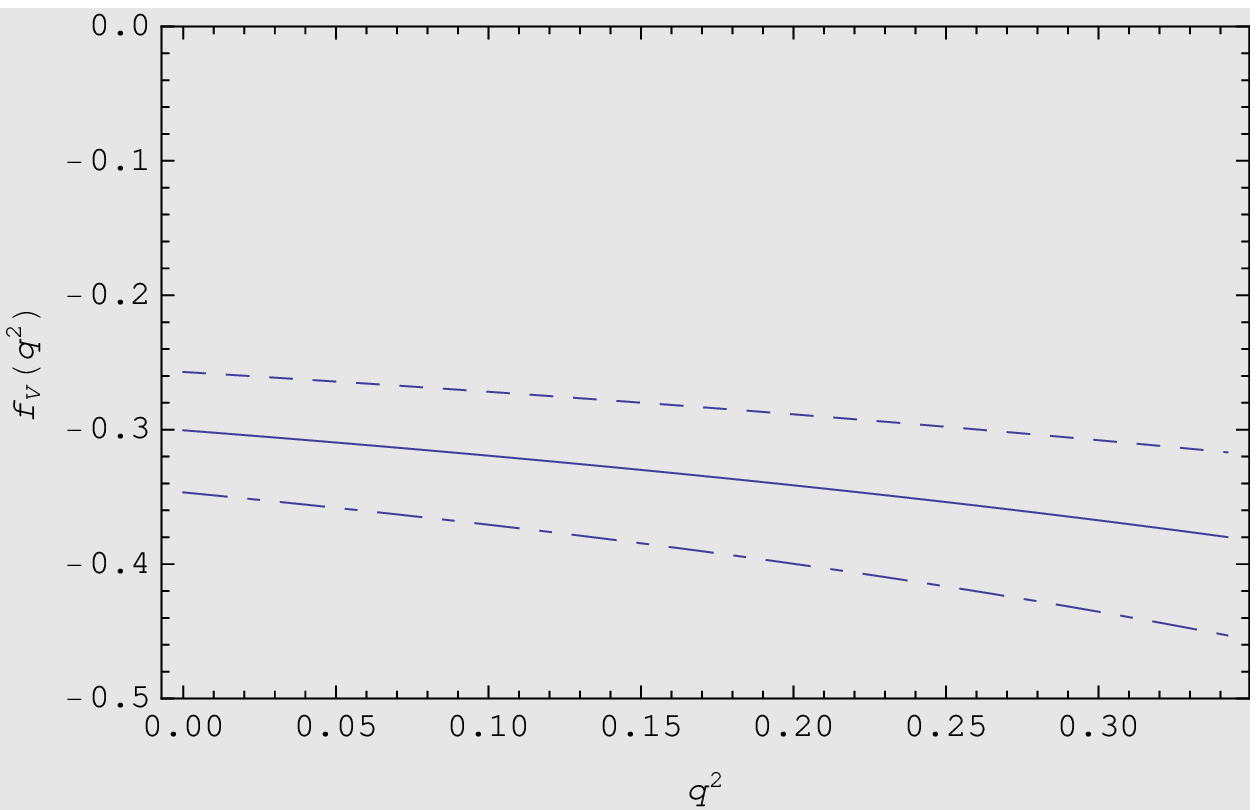}\\
\caption{ \label{FLqdpf1} The variation of $D^+ \rightarrow
f_1(1285)$ transition form factors as functions of $q^2$. The solid,
dot dashed and dashed lines correspond to the center value, upper
and lower limit respectively.}
\end{figure}

\begin{figure}[htbp]
\includegraphics[width=6cm]{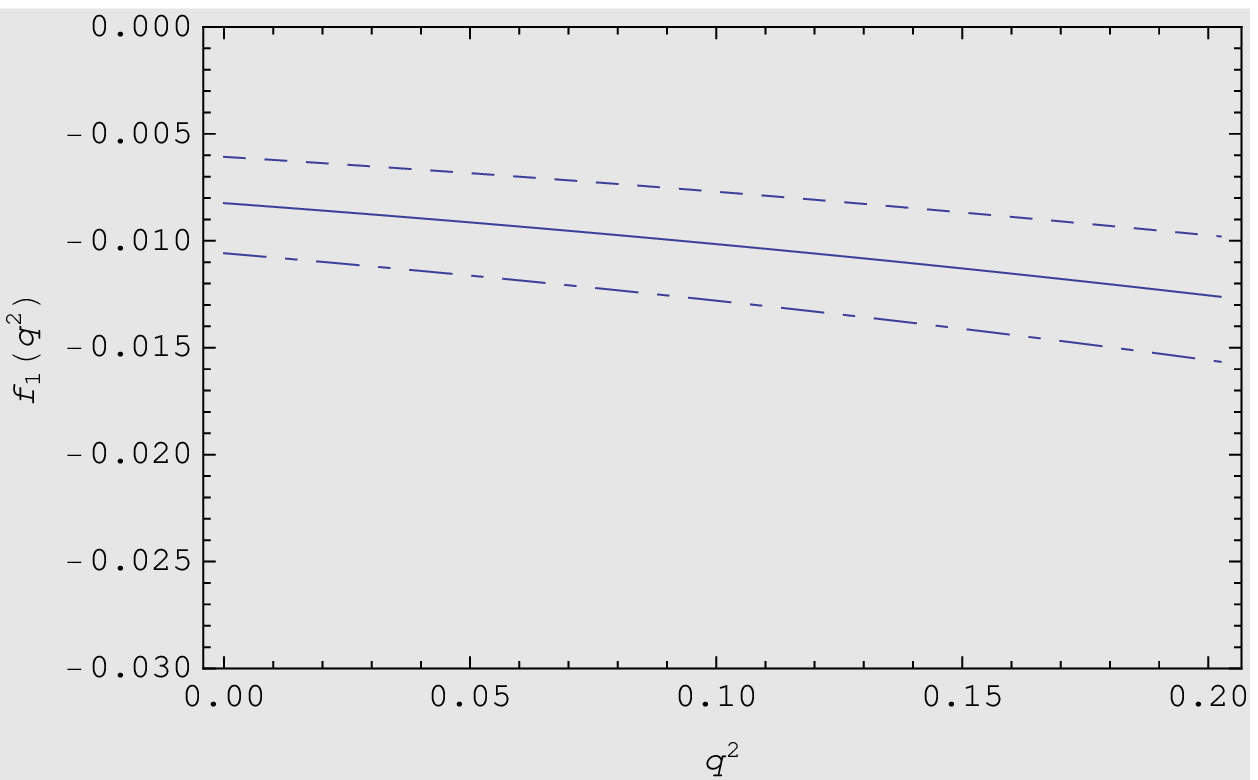}\hfill
\includegraphics[width=6cm]{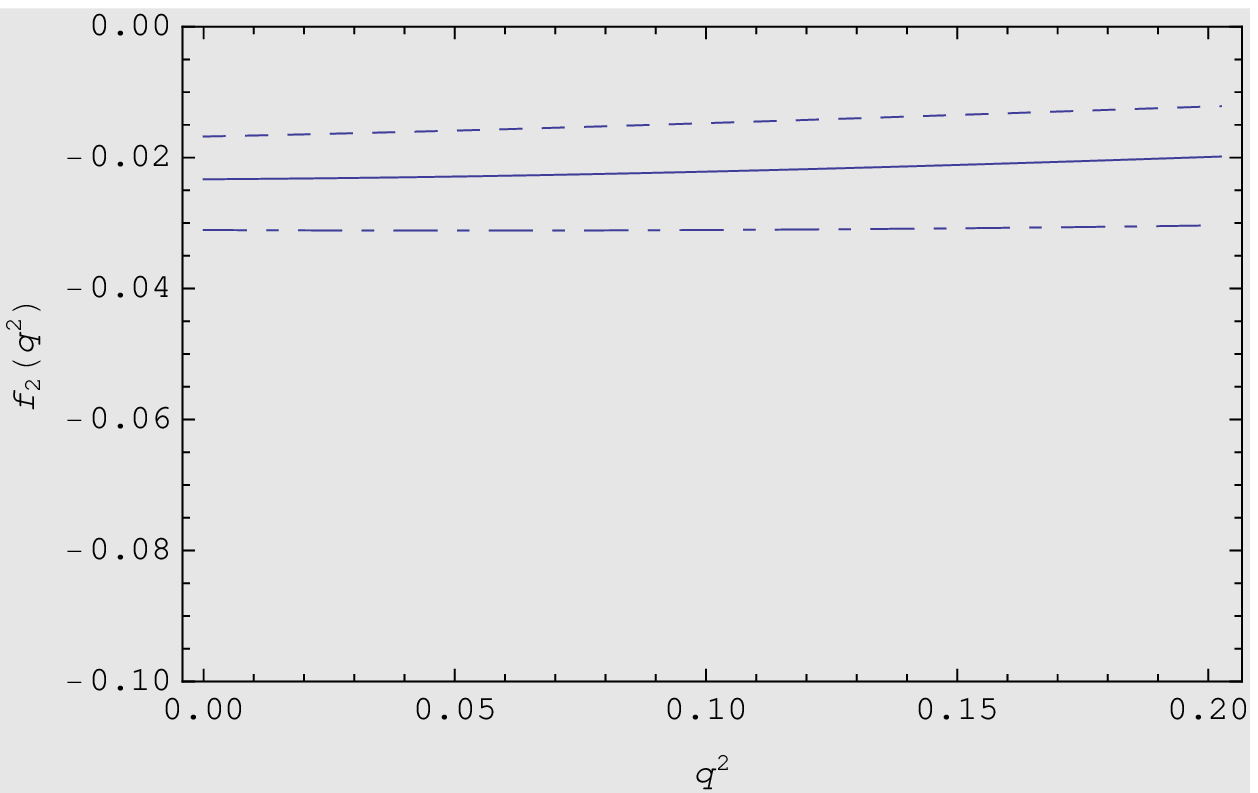}\\
\includegraphics[width=6cm]{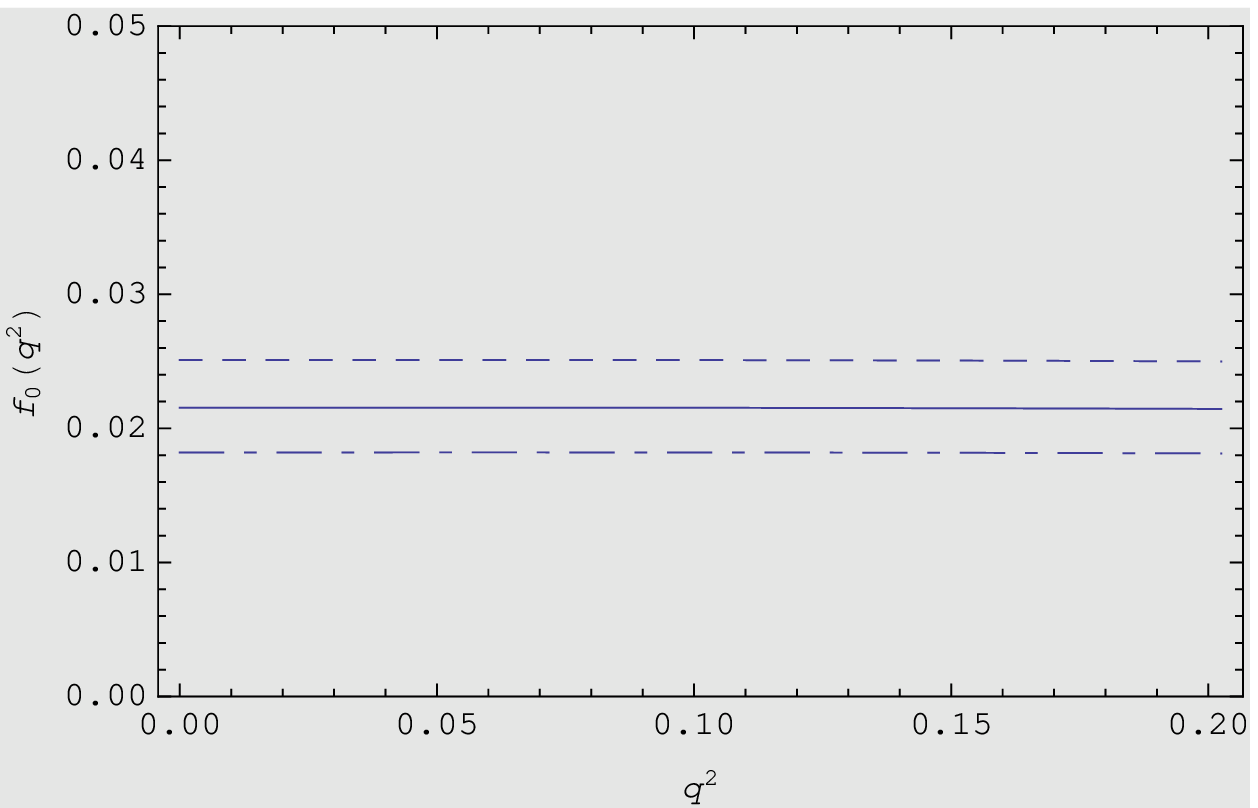}\hfill
\includegraphics[width=6cm]{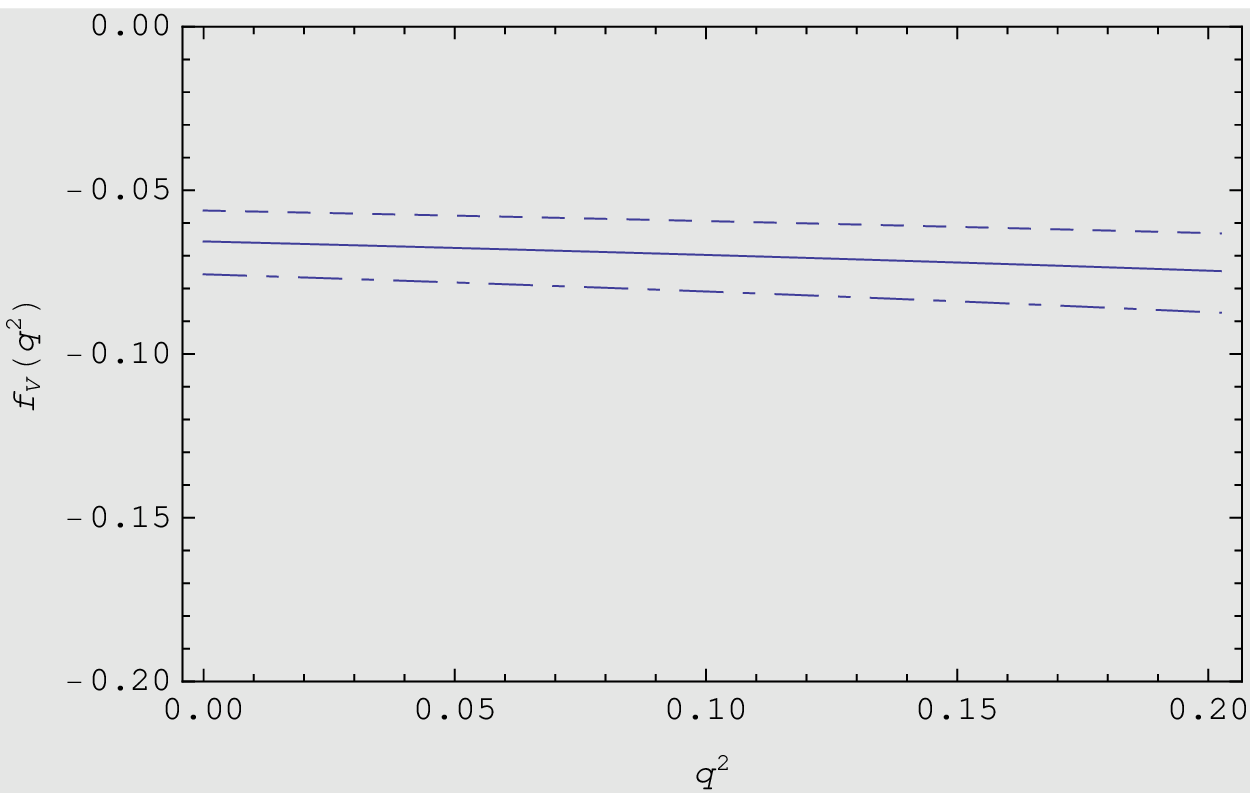}\\
\caption{ \label{FHqdpf1} The variation of $D^+ \rightarrow
f_1(1420)$ transition form factors as functions of $q^2$. The solid,
dot dashed and dashed lines correspond to the center value, upper
and lower limit respectively.}
\end{figure}

It is easily seen from Fig.\ref{FM1dpa1}-\ref{FHM2dpf1} that the
form factors $f_1$ and $f_V$ are more insensitive to the
$s^\prime_0$ and $s_0$ respectively. Concretely, we choose the free
parameters $s_0 = 7.0 \pm 0.2 {\rm GeV^2}$, $s_0^\prime = 3.5 \pm
0.2 {\rm GeV^2}$, $M^2_1 = 8.0 \pm 0.5 {\rm GeV^2}$, $M^2_2 = 7.0
\pm 0.5 {\rm GeV^2}$ in our calculations. Note that the sum rules
become unreliable in the large $q^2$ region and therefore our
predictions for the form factors via 3-point sum rules are truncated
at about $q^2=0.15 {\rm GeV}^2$. To extend the results to the whole
kinematically allowed region, i.e. $0 \leq q^2 \leq ( m_D-m_A)^2$,
we use the following parametrization,
\begin{eqnarray}
F (q^2) = \frac{F(0)}{ 1- a (q^2/m^2_D) + b ( q^2/m^2_D)^2}
\end{eqnarray}
Here $F$ denotes the form factor $f_1,f_2,f_0,f_V$. By using this
parametrization and the form factors at low $q^2$ region calculated
via 3-point sum rules, we obtain the form factors in the whole
kinematically allowed region shown in
Fig.\ref{Fqdpa1}-\ref{FHqdpf1}. The $D^+ \rightarrow a_1^0(1260),
D^0 \rightarrow a_1^-(1260), D^+ \rightarrow f_1 (1285),f_1(1420)$
transition form factors at $q^2=0$ and corresponding extrapolative
parameters $a,b$ are exhibited in Table \ref{tab2}, where the
theoretical errors for $a, b$ are not shown for simplicity.

\begin{table}[htbp]
\tbl{ The $D^+ \rightarrow a_1^0(1260)$, $D^0 \rightarrow
a_1^-(1260)$, $D^+ \rightarrow f_1(1285), f_1(1420)$ transition form
factors at $q^2=0$ and corresponding extrapolative parameters
$a,b$.}
{\begin{tabular*}{10cm}{@{}ccccc@{}} \toprule Decay & \multicolumn{2}{c}{$F(0)$}   & a & b\\
\colrule
$D^+ \rightarrow a_1^0(1260)$ & $f_1$ & $0.039^{+0.012}_{-0.010}$ & 7.18 & 20.60 \\
 & $f_2$ & $0.112^{+0.037}_{-0.032}$ & -0.68 & 40.47 \\
& $f_0$ & $-0.114^{+0.018}_{-0.019} $ & 0.08 & 2.47 \\
& $f_V $ & $0.314^{+0.048}_{-0.046} $ & 2.04 & -1.00 \\
\hline
$D^0 \rightarrow a_1^-(1260)$ & $f_1$ & $-0.057^{+0.015}_{-0.016}$ & 7.04 & 19.65 \\
 & $f_2$ & $-0.154^{+0.045}_{-0.052}$ & -0.72 & 42.52 \\
& $f_0$ & $0.161^{+0.027}_{-0.025} $ & 0.08 & 2.49 \\
& $f_V $ & $-0.442^{+0.064}_{-0.068} $ & 2.03 & -0.98 \\
\hline
$D^+ \rightarrow f_1(1285)$ & $f_1$ & $-0.038^{+0.010}_{-0.010}$ & 7.18 & 20.60 \\
&  $f_2$ & $-0.107^{+0.030}_{-0.035}$ & -0.68 & 40.47 \\
& $f_0$ & $0.107^{+0.018}_{-0.016} $ & 0.08 & 2.47 \\
& $f_V $ & $-0.300^{+0.043}_{-0.047} $ & 2.04 & -1.00 \\
\hline
$D^+ \rightarrow f_1(1420)$ & $f_1$ & $-0.008^{+0.002}_{-0.002}$ & 7.18 & 20.60 \\
& $f_2$ & $-0.023^{+0.007}_{-0.008}$ & -0.68 & 40.47 \\
& $f_0$ & $0.022^{+0.004}_{-0.003} $ & 0.08 & 2.47 \\
& $f_V $ & $-0.066^{+0.009}_{-0.010} $ & 2.04 & -1.00 \\
 \botrule
\end{tabular*} \label{tab2}}
\end{table}



Roughly speaking, the uncertainties of the form factors $f_1$, $f_0$
and $f_V$ are about (15-20)\%, while that of $f_2$ is around 30\%
which can be attributed to the large dependence of this form factor
on the free parameters $s_0, s_0^\prime, M^2_1, M^2_2$. In addition,
the $D^+ \rightarrow a_1^0 ( 1260)$ transition form factors have
opposite signs compared with those of other channels. This is
because these decays are governed by the $c \rightarrow d$
transition in the quark level and the flavor content of $a_1^0
(1260)$ is $1/\sqrt{2}(u\bar{u} - d \bar{d})$ where the sign of $d
\bar{d}$ component is minus. Moreover, the magnitude of form factor
$f_1$ is obviously smaller than other form factors and the form
factor $f_0$ are almost independent of $q^2$ for all channels we
investigated. For the $D^+ \rightarrow f_1 (1285), f_1(1420)$ decays
in which mixing is involved, the form factors of $D^+ \rightarrow
f_1 (1285)$ are about five times larger than those of $D^+
\rightarrow f_1 (1420)$, which is due to the fact that the flavor
contents of $f_1 (1285)$ and $f_1 (1420)$ are dominated by the
components $1/\sqrt{2}(u \bar{u} + d \bar{d})$ and $s \bar{s}$
respectively. The extrapolative parameters $a, b$ are the same for
$D^+ \rightarrow a_1^0(1260), f_1(1285), f_1(1420)$ decays and this
means the variations of corresponding transition form factors in
these three channels with the momentum transfer squared $q^2$ are
similar.

\section{Branching ratios for semileptonic decays}

In this section, we shall use the above obtained transition form
factors to calculate the branching ratios of relevant semileptonic
decays. Specially, the decay channels we calculate are $D^+
\rightarrow a_1^0 (1260) l^+ \nu_l, D^0 \rightarrow a_1^-(1260) l^+
\nu_l, D^+ \rightarrow f_1(1285) l^+ \nu_l$ and $D^+ \rightarrow f_1
(1420) l^+ \nu_l$. The lepton $l$ may be an electron or a muon.

The differential decay widths for the process $D \rightarrow A l^+
\nu_l$ can be written as
\begin{eqnarray}
\frac{d \Gamma}{d q^2} = \frac{d \Gamma_T}{d q^2} + \frac{d
\Gamma_L}{d q^2}
\end{eqnarray}
where $\frac{d \Gamma_T}{d q^2}$ and $\frac{d \Gamma_T}{d q^2}$
denote the transverse and longitudinal differential decay width
respectively. Neglecting the mass of lepton $l$, their expressions
have the following forms,
\begin{eqnarray}\label{DWT}
\frac{d \Gamma_T}{d q^2} = \frac{G_F^2 |V_{cd}|^2}{192 \pi^3 m^3_D}
q^2 \lambda^{1/2} ( m^2_D, m^2_A, q^2)  ( |H_+|^2+ |H_-|^2 )
\end{eqnarray}
\begin{eqnarray}
\frac{d \Gamma_L}{d q^2} = \frac{G_F^2 |V_{cd}|^2}{192 \pi^3 m^3_D}
\lambda^{1/2} ( m^2_D, m^2_A, q^2) |H_0|^2
\end{eqnarray}
with
\begin{eqnarray}
& & H_\pm (q^2) = (m_D +m_K) f_0(q^2)  \mp \frac{\lambda^{1/2}
(m^2_D,
m^2_A, q^2)}{m_D + m_A} f_V (q^2) \nonumber \\
& & H_0 (q^2) = \frac{1}{ 2 m_A } \left [ ( m^2_D - m^2_A -q^2 ) (
m_D + m_A) f_0 (q^2) \right. \nonumber \\
& & \hspace{1.5cm} \left. - \frac{\lambda^{1/2} (
m^2_D, m^2_A, q^2)}{ m_D + m_A } f_1 ( q^2) \right ]   \nonumber \\
& & \lambda (m^2_D, m^2_A, q^2)= m^2_D + m^2_A + q^4 - 2m^2_D q^2
 -2 m^2_A q^2 - 2 m^2_D m^2_A \nonumber
\end{eqnarray}

\begin{figure}
\includegraphics[width=6cm]{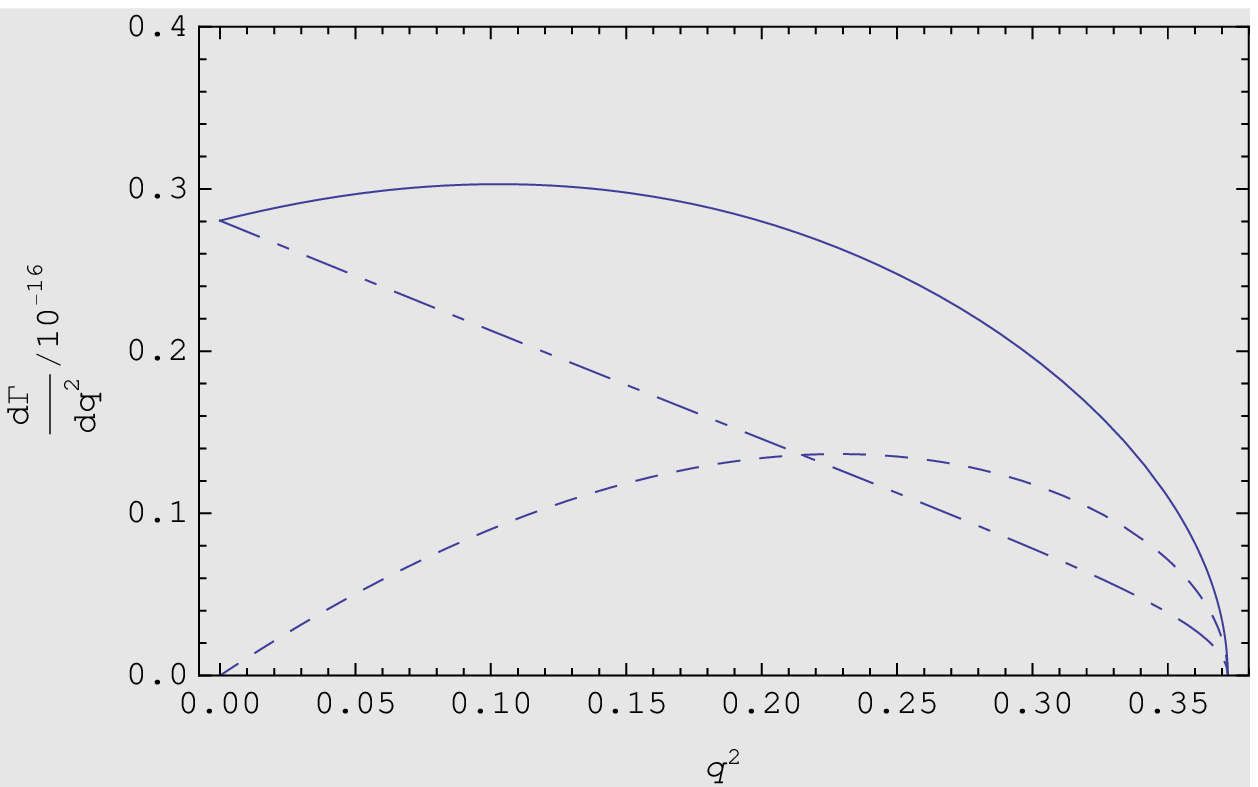}\hfill
\includegraphics[width=6cm]{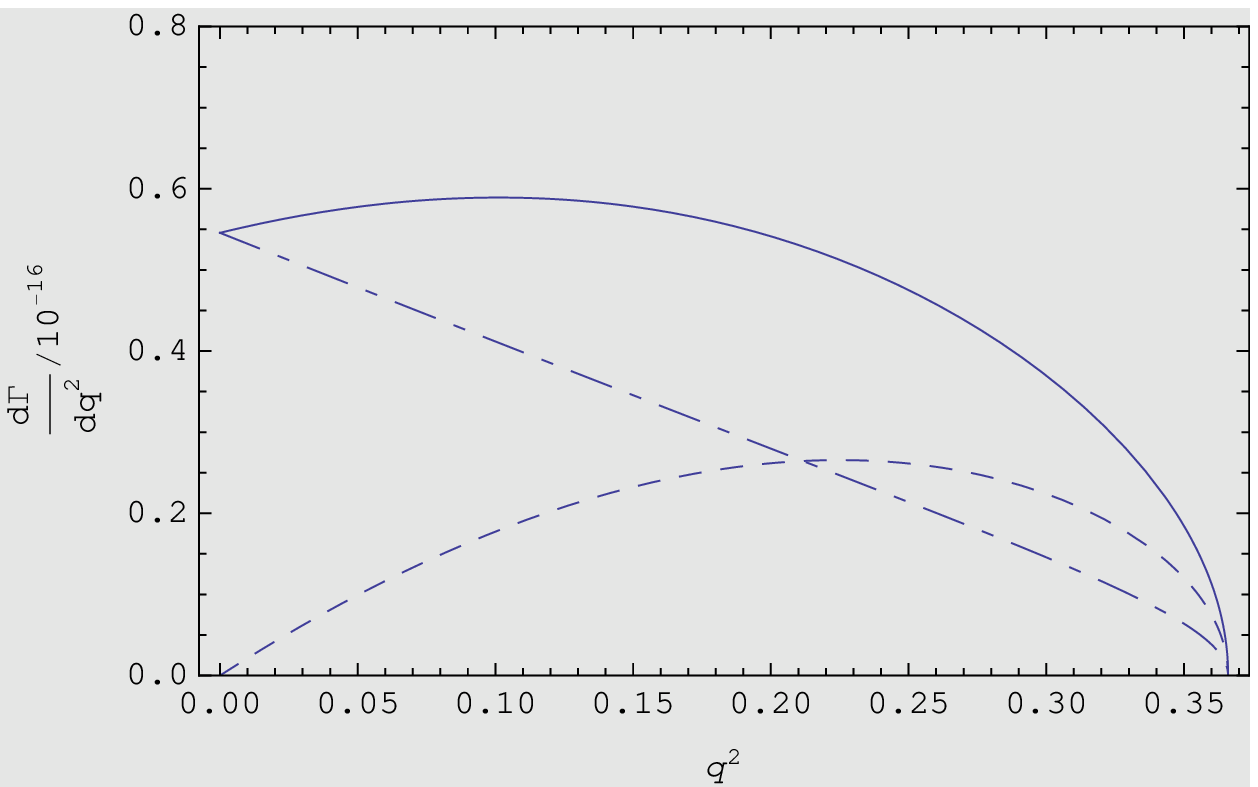}\\
\includegraphics[width=6cm]{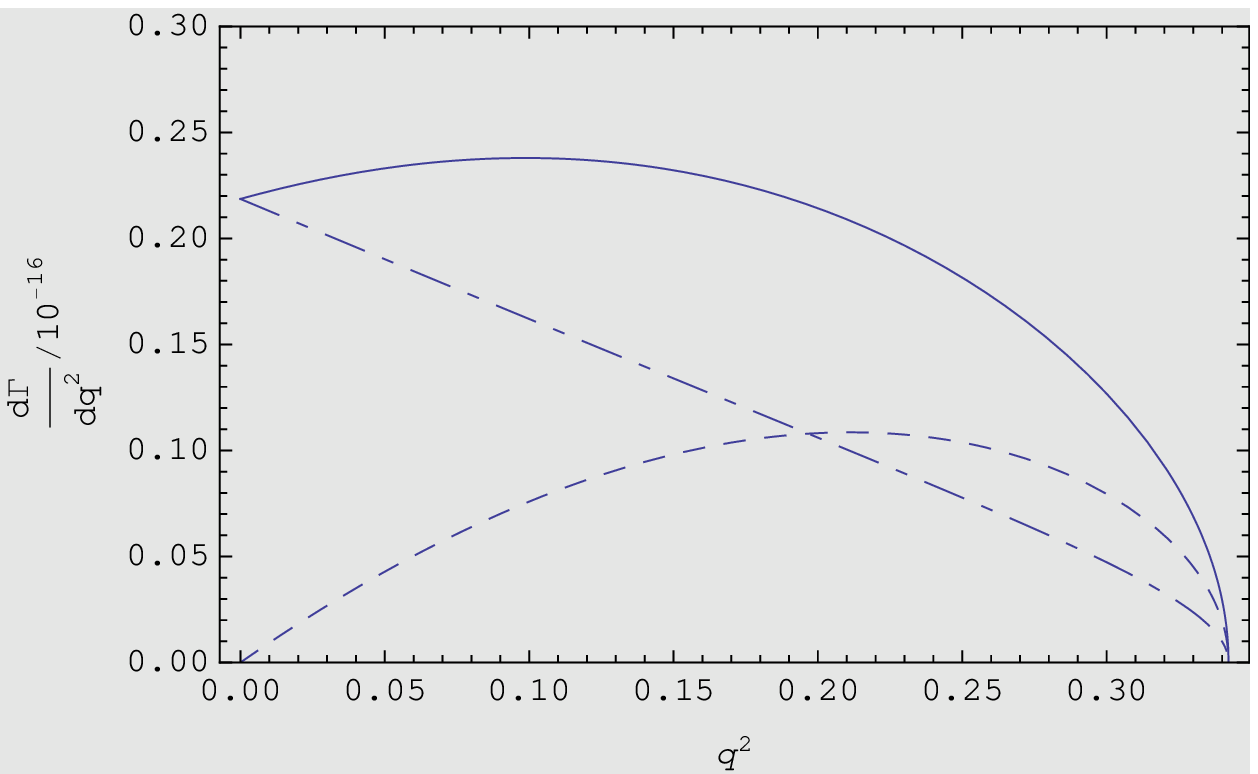}\hfill
\includegraphics[width=6cm]{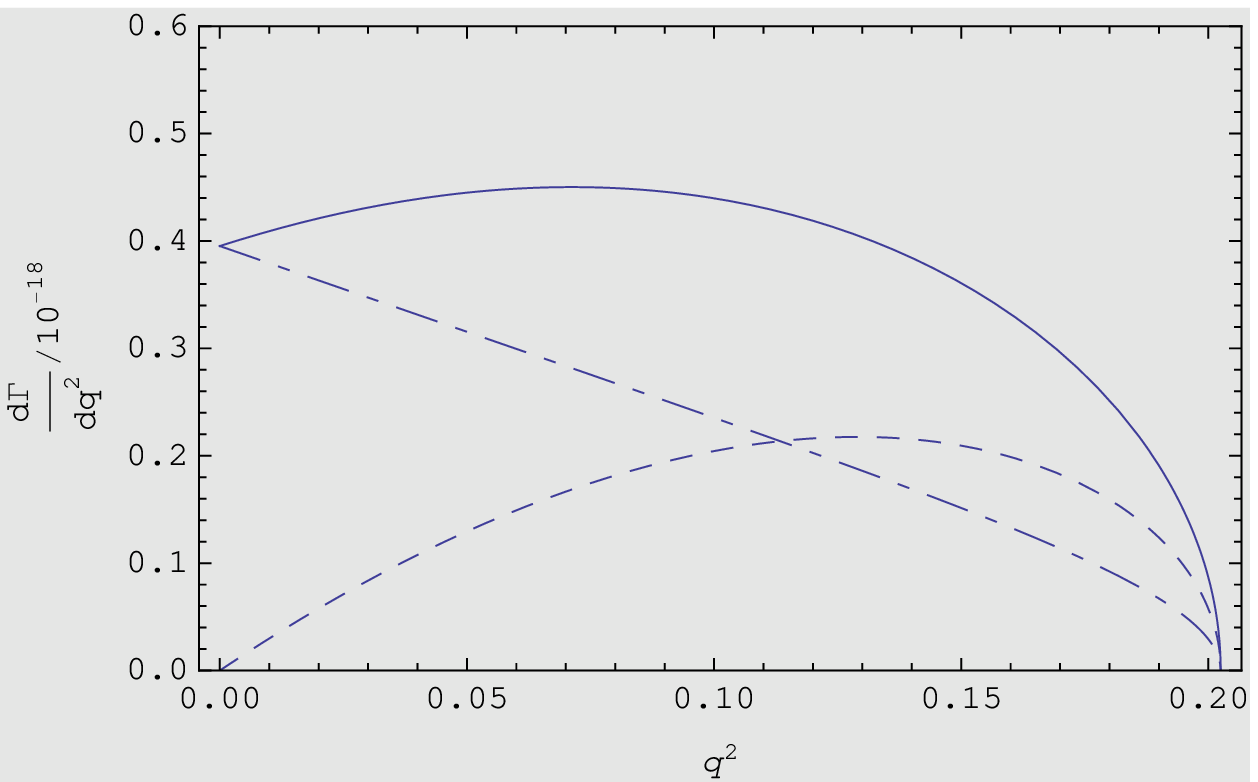}
\caption{The differential decay widths of $ D^+ \rightarrow
a_1^0(1260) l^+ \nu_l$, $D^0 \rightarrow a_1^- (1260) l^+ \nu_l$,
$D^+ \rightarrow f_1(1285)l^+ \nu_l$ and $D^+ \rightarrow f_1(1420)
l^+ \nu_l$ as functions of $q^2$, which are shown in the upper-left,
upper-right, lower-left and lower-right plots respectively. The
dashed, dot dashed and solid lines correspond to the transverse,
longitudinal and total differential decay width
respectively.}\label{drates}
\end{figure}

For the CKM matrix element $|V_{cd}|$ and Fermi coupling constant
$G_F$, we use the latest values given by PDG, i.e. $|V_{cd}|= 0.225
\pm 0.008$, $G_F = 1.116 \times 10^{-5} {\rm GeV}^{-2}$.  With the
above considerations, we obtain the differential decay widths as
functions of $q^2$ for $D^+ \rightarrow a_1^0 (1260) l^+ \nu_l, D^0
\rightarrow a_1^-(1260) l^+ \nu_l, D^+ \rightarrow f_1(1285) l^+
\nu_l$ and $D^+ \rightarrow f_1 (1420) l^+ \nu_l$ decays, the center
values of which are shown in Fig.\ref{drates}.

From Fig.\ref{drates}, we can observe that the differential decay
widths for $D^+ \rightarrow a_1^0 (1260) l^+ \nu_l$ are about half
of those for $D^0 \rightarrow a_1^- l^+ \nu_l$ which can be ascribed
to the factor $1/\sqrt{2}$ in the flavor content of $a_1^0$(see
Eq.(\ref{flavor})). In addition, the differential decay widths for
$D^+ \rightarrow f_1 (1420) l^+ \nu_l $ are almost two order of
magnitude smaller than those for $D^+ \rightarrow f_1 (1285) l^+
\nu_l $ and this is related to the fact that $f_1(1420)$ is nearly a
pure $s \bar{s}$ state. The transverse differential decay width
vanishes at the largest recoil, i.e. $q^2=0$ point due to the factor
$q^2$ in the front of Eq.(\ref{DWT}).

Integrating the differential decay widths over $q^2$ and using the
lifetimes of $D$ mesons given by PDG, i.e. $\tau_{D^+} = 1.04 \times
10^{-12} s$, $\tau_{D^0} = 0.41 \times 10^{-12} s$, we get the
following branching ratios for relevant semileptonic decays,
\begin{eqnarray}
 Br ( D^+ \rightarrow a_1^0(1260) l^+ \nu_l ) =
1.47^{+0.44+0.11}_{-0.34-0.10} \times 10^{-5} \\
 Br ( D^0 \rightarrow a_1^- (1260) l^+ \nu_l ) =
1.11^{+0.33+0.08}_{-0.26-0.08} \times 10^{-5} \\
 Br ( D^+ \rightarrow f_1(1285) l^+ \nu_l ) =
1.07^{+0.32+0.07}_{-0.25-0.08} \times 10^{-5} \\
 Br ( D^+ \rightarrow f_1 (1420) l^+ \nu_l ) =
1.22^{+0.39+0.09}_{-0.31-0.09} \times 10^{-7}
\end{eqnarray}
where the first  and second uncertainties come from the transition
form factors and CKM matrix element $|V_{cd}|$ respectively. The
uncertainties coming from transition form factors are about
(20-30)\% and those stemming from $|V_{cd}|$ are around 7\%. The
branching ratios for $D^+ \rightarrow a_1^0 (1260) l^+ \nu, D^0
\rightarrow a_1^-(1260) l^+ \nu_l$ and $D^+ \rightarrow f_1 (1285)
l^+ \nu$ are at the order of ${\cal O} ( 10^{-5} )$, while the one
for $D^+ \rightarrow f_1 (1420) l^+ \nu$ is approximately two order
of magnitude smaller which is similar to the differential decay
width. Note that these branching ratios have not been measured, our
results can be tested by the more precise experiments such as LHCb,
BelleII,etc. in the future.

\section{Conclusion}

In the present paper, we calculate the $D$ to spin triplet axial
vector meson $a_1^0 (1260)$, $a_1^- (1260)$, $f_1(1285)$,
$f_1(1420)$ transition form factors by using the 3-point QCD sum
rules. The flavor contents of each meson and the mixing between $f_1
(1285)$ and $f_1 (1420)$ are considered in detail. The uncertainties
of transition form factors comes from the four free parameters $s_0,
s_0^\prime, M^2_1, M^2_2$ which is about (15-20)\% for $f_1, f_0,
f_V$ and 30\% for $f_2$. The transition form factors of $D^+
\rightarrow f_1(1285)$ are about five times larger than those of
$D^+ \rightarrow f_1 (1420)$ due to the fact that the flavor
contents of $f_1(1285)$ and $f_1 (1420)$ are dominated by the
components $1/\sqrt{2} ( u \bar{u} + d \bar{d} )$ and $s \bar{s}$
respectively. Based on the form factors obtained here, we predict
the branching ratios of relevant semileptonic decays. The branching
ratios for $D^+ \rightarrow a_1^0 (1260) l^+ \nu, D^0 \rightarrow
a_1^- l^+ \nu_l$ and $D^+ \rightarrow f_1 (1285) l^+ \nu$ are at the
order of ${\cal O} ( 10^{-5} )$, while the one for $D^+ \rightarrow
f_1 (1420) l^+ \nu$ is approximately two order of magnitude smaller.
These results can be tested by the more precise experiments in the
future.

\section*{Acknowledgments}

This research is supported by the Ph.D Programs Foundation of
Ministry of Education of China under Grant No.20132136120003.





\end{document}